\documentclass[12pt]{article}
\pdfoutput=1
\usepackage{putex}
\usepackage{amsmath,hyperref,amssymb,amsfonts,bm,amscd,slashed,ytableau}
\usepackage{cite}
\usepackage{graphicx}
\usepackage{multirow}
\usepackage{verbatim}
\usepackage{appendix}
\usepackage{fancybox}
\usepackage{array}
\usepackage{url}
\usepackage{float}
\usepackage{mathtools}
\usepackage{mathrsfs}
\usepackage{bbm}
\usepackage[bbgreekl]{mathbbol}

\DeclareSymbolFontAlphabet{\mathbbm}{bbold}
\DeclareSymbolFontAlphabet{\mathbb}{AMSb}

\DeclareMathAlphabet{\mathpzc}{OT1}{pzc}{m}{it}

\newcommand{\bea}{\begin{eqnarray}}
\newcommand{\eea}{\end{eqnarray}}
\newcommand{\be}{\begin{equation}}
\newcommand{\ee}{\end{equation}}

\def \beaa {\begin{equation}\begin{aligned}}
\def \eeaa {\end{aligned}\end{equation}}

\newcommand{\Z}{{\mathbb Z}}
\newcommand{\R}{{\mathbb R}}
\newcommand{\C}{{\mathbb C}}

\newcommand{\cF}{{\mathcal{F}}}
\newcommand{\cN}{{\mathcal{N}}}

\newcommand{\dd}{{\rm d}}

\def\Tr{{\rm Tr \,}}

\def\bar{\overline}

\def\cA{{\mathcal A}}

\def\cF{{\mathcal F}}

\def\cI{{\mathcal I}}

\def\cL{{\mathcal L}}
\def\cM{{\mathcal M}}
\def\cN{{\mathcal N}}
\def\cO{{\mathcal O}}

\def\cR{{\mathcal R}}

\def\cT{{\mathcal T}}

\def\cV{{\mathcal V}}
\def\cW{{\mathcal W}}

\def\cY{{\mathcal Y}}

\renewcommand{\bar}{\overline}

\numberwithin{equation}{section}
\begin{document}

\institution{SCGP}{Simons Center for Geometry and Physics,\cr Stony Brook University, Stony Brook, NY 11794-3636, USA}
\institution{Perimeter}{Perimeter Institute for Theoretical Physics, Waterloo, ON N2L 2Y5, Canada}

\title{Correlators on the wall and $\mathfrak{sl}_n$ spin chain}
\authors{Mykola Dedushenko\worksat{\SCGP} and Davide Gaiotto\worksat{\Perimeter}}

\abstract{We study algebras and correlation functions of local operators at half-BPS interfaces engineered by the stacks of D5 or NS5 branes in the 4d $\cN=4$ super Yang-Mills. The operator algebra in this sector is isomorphic to a truncation of the Yangian $\mathcal{Y}(\mathfrak{gl}_n)$. The correlators, encoded in a trace on the Yangian, are controlled by the inhomogeneous $\mathfrak{sl}_n$ spin chain, where $n$ is the number of fivebranes: they are given in terms of matrix elements of transfer matrices associated to Verma modules, or equivalently of products of Baxter's Q-operators. This can be viewed as a novel connection between the $\cN=4$ super Yang-Mills and integrable spin chains. We also remark on analogous constructions involving half-BPS Wilson lines.}

\date{}

\maketitle

\tableofcontents

\section{Introduction and Conclusions}
This paper is a second part of work on supersymmetric local observables on boundaries and interfaces in the maximal super Yang-Mills (MSYM) in four dimensions \cite{DG1}. Here our focus is on interfaces engineered by fivebranes intersecting stacks of D3 branes along three-dimensional loci \cite{Gaiotto:2008sa,Gaiotto:2008ak}. We identify algebras equipped with twisted traces that encode the data of correlators in the topological quantum mechanics (TQM) subsector on the interface. The algebras are given by certain truncations of the Yangians \cite{Ishtiaque:2018str}, and the correlation functions are computed in terms of transfer matrices of an inhomogeneous $\mathfrak{sl}_n$ spin chain.

We consider both interfaces engineered by the D5 branes and their S-dual NS5 branes. The basic setting is $N$ D3 branes crossing a stack of $n$ fivebranes, with possible $n\,s$ D3 branes terminating from the right. We denote the corresponding algebras as $\cA^{(s)}_{N;n}$, or simply $\cA_{N;n}$ if $s=0$. When $s>0$, we only consider the configurations with exactly $s$ D3 branes terminate on each fivebrane, so the $U(n)$ symmetry on the fivebranes is unbroken. This ultimately leads to the morphism from the Yangian $\cY[\mathfrak{gl}_n]$ to the algebra $\cA^{(s)}_{N;n}$. We do not discuss generalizations to configurations with different numbers of D3 branes terminating on each fivebrane, but they are straightforward and lead to morphisms from shifted Yangians \cite{Braverman:2016pwk}.

The results presented here can be viewed as a new connection between supersymmetric QFT (SQFT) and integrability, where for the SQFT we mostly take the 4d MSYM. It is different from the well-known Yangian symmetry $\cY[\mathfrak{psu}_{2,2|4}]$ of the MSYM \cite{Minahan:2002ve,Beisert:2003tq,Beisert:2003yb,Dolan:2004ps,Drummond:2008vq,Drummond:2009fd}, for we are only concerned with the supersymmetric sector of the theory, i.e. the appropriate Q-cohomology. Furthermore, it is not limited to 4d MSYM, as some of our results are about 3d $\cN=4$ gauge theories. In this respect, our findings are more in the spirit of Bethe/Gauge correspondence \cite{Nekrasov:2009uh,Nekrasov:2009ui,Nekrasov:2009rc,Nekrasov:2011bc,Nekrasov:2014xaa}, which deals with integrability in the supersymmetric sectors of theories with 8 supercharges. The apparent similarity is closer for interfaces built from a stack of $n$ NS5 branes: they are described by three-dimensional $A_{n-1}$ quiver theories, which are related to the $\mathfrak{sl}_n$ spin chain via the Bethe/Gauge correspondence. Our results also relate them to the $\mathfrak{sl}_n$ spin chain, but there are crucial differences: for \cite{Nekrasov:2009uh}, the spin chain length $L$ is determined by the number of flavors, while for us $L$ is fixed by the number of insertions in the correlator of the 3d theory. For us, the number of flavors is related to the choice of a collection of infinite-dimensional modules of the Yangian that determine the twisted trace. Incidentally, in \cite{Nekrasov:2009uh}, the number of flavors determines the finite-dimensional module of the Yangian (the spin chain Hilbert space). Thus a unifying theme of the two approaches is the relation of quantum field theories to the representation theory of the Yangian: finite-dimensional in \cite{Nekrasov:2009uh}, and infinite-dimensional for us. Our main example in the second half of this paper is the $A_1$ quiver related to the XXX spin chain, for which the standard Bethe/Gauge correspondence is studied in \cite{LeeNek}.

\begin{figure}[t]
	\label{fig:s3s4}
	\centering
	\includegraphics[scale=0.9]{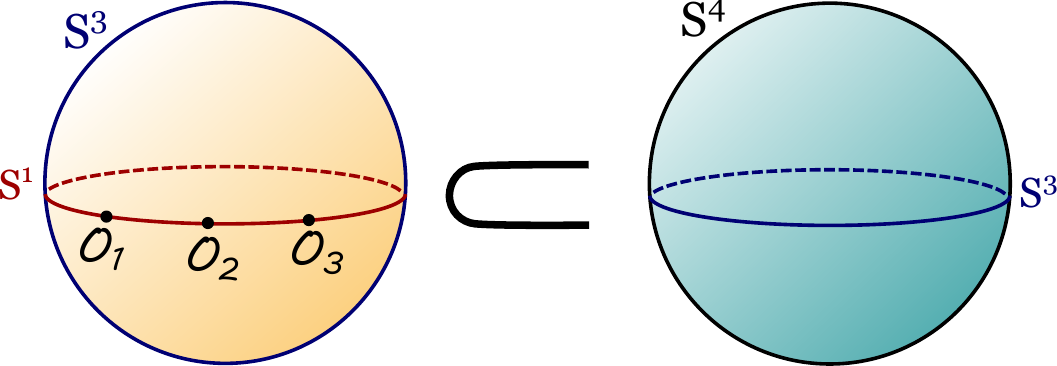}
	\caption{Local operators live on $S^1\subset S^3$, and $S^3$ is the equator of $S^4$.}
\end{figure}

General constructions of the boundary or interface TQM are detailed in \cite{DG1}, so we do not review them and move directly to the applications. We only give a sketch in Figure \ref{fig:s3s4}, and note that those constructions are based on protected 1d sectors in 3d $\cN=4$ theories introduced in \cite{Chester:2014mea,Beem:2016cbd}. One can construct algebras either using the superconformal definition, or the Omega background, or the sphere background. To compute correlators, it is important to use the sphere background \cite{Dedushenko:2016jxl}, see Figure \ref{fig:s3s4}. Results for certain observables in the same setup have appeared recently in \cite{Wang:2020seq,Komatsu:2020sup}: indeed, they look at the same protected sector in 4d MSYM with an interface. Nonetheless, there is no significant overlap with those works as we emphasize the algebraic approach and study different correlation functions. Also note that the boundary/interface algebras describe short quantizations of Poisson varieties \cite{Etingof:2019guc}.

The six-dimensional super Yang-Mills (SYM) on the D5 branes provides an alternative viewpoint on our constructions, which we do not take in most of this work and only briefly discuss now. On the D5 brane worldvolume, the couplings with D3 branes engineer codimension-3 defects in the 6d $U(n)$ SYM, which are 't Hooft operators. As the brane system is embedded in IIB string theory, these couplings must be compatible 
with a coupling to supergravity. In particular, they will be compatible with the $\Omega$ deformation in a plane shared by all the branes. Truncation to the SUSY sector in this background reduces the 6d theory to the holomorphic-topological 4d Chern-Simons (CS): see \cite{Costello:2013zra,Costello:2013sla,Costello:2017dso,Costello:2018gyb} for explorations of the 4d CS, \cite{NikiThesis} for an early proposal, and \cite{Costello:2018txb} on how it descends from 6d. Along the way, the defect becomes the usual 't Hooft line in 4d, which is a subject of investigation in \cite{CGY}. Consistency of the coupling to the defect
requires the existence of an algebra morphism from the Yangian algebra $\cY[\mathfrak{gl}_n]$ to ${\cal A}_{N;n}$ or $\cA^{(s)}_{N;n}$.
Essentially, the morphism identifies which 1d operators are coupled to the holomorphic derivatives of the 4d CS connection. Our first objective is to identify this morphism, and we will indeed find that all three classes of algebras, $\cA_{N;n}$, $\cA^{(1)}_{N;n}$, and $\cA^{(s)}_{N;n}$, admit homomorphisms from the Yangian $\cY[\mathfrak{gl}_n]$. The second objective is to compute traces on these algebras that encode defect correlators of the 4d SYM, which is in fact central to this paper.

Finally, there is a close relation to the results in \cite{Giombi:2018qox,Giombi:2020amn}, who also studied topological correlators in 4d MSYM, however they look at local observables on Wilson loops. Indeed, Wilson loops associated to D5 branes are also associated to Yangian embeddings in protected subsectors, 
which can be explained by the same $\Omega$ deformation of 6d SYM argument. See also \cite{Ishtiaque:2018str}. Interfaces and Wilson loops can even be combined to give a setup controlled by the $\cY[\mathfrak{gl}_{n|m}]$ Yangian. We will discuss this very briefly in the main text in Section \ref{wilson}.

In Section \ref{sec:Higgs_alg} we study algebras living at the interfaces engineered by D5 branes. Using the techniques developed in \cite{DG1}, we identify $\cA_{N;n}$, i.e. the algebra at the intersection of $N$ D3 and $n$ D5 branes, as a quantum Hamiltonian reduction of a tensor product of two copies of $U(\mathfrak{gl}_N)$ and $Nn$ copies of the Weyl algebra. We also describe the morphism from $\cY[\mathfrak{gl}_n]$ very explicitly in the RTT presentation of the Yangian \cite{Baxter:1972hz,Takhtajan:1979iv}. We then construct the algebra $\cA^{(1)}_{N;n}$ as the $\mathfrak{gl}_N$-invariant subalgebra of $U(\mathfrak{gl}_{N+n})$, with its known maps from the Yangian. At the end, we apply the theory of finite W-algebras to identify $\cA^{(s)}_{N;n}$ and the corresponding map from $\cY[\mathfrak{gl}_n]$ in the RTT presentation.

In Section \ref{sec:Coulomb} we describe algebras at the NS5-like interfaces. Unlike in the D5 case, it does not require separate treatment for the $s=0$, $s=1$ and $s>1$ cases. The algebra can be described uniformly for all $s$ as the Coulomb branch algebra of a certain balanced linear quiver, which is then coupled to the 4d MSYM. It comes equipped with a certain representation in terms of shift operators, which allows to define the twisted trace and correlators. The map from the Yangian is also given explicitly, and the representation in terms of shift operators, from the Yangian point of view, is the well-known GKLO representation \cite{Gerasimov_2005}. The Coulomb branch perspective, notably, leads to the Drinfeld's second presentation of the Yangian \cite{Drinfeld:1987sy}. The Coulomb branch algebra itself is the quotient of the Yangian \cite{Braverman:2016pwk} known as a truncated Yangian \cite{Cher,BRUNDAN2006136} (or more generally shifted truncated Yangian in the unbalanced case). It admits a family of ``coproduct'' maps that are compatible with the usual Yangian coproduct, as has been proven in \cite[Proposition 4.1.13]{WeekesThesis} in the $A_n$ case (see also \cite{FINKELBERG2018349}). For a shifted and truncated Yangian $\cY_\nu^\lambda[\mathfrak{sl}_n]$, both shift and truncation parameters split under the coproduct, $\nu=\nu_1+\nu_2$, $\lambda=\lambda_1+\lambda_2$, so that we get $\cY_\nu^\lambda[\mathfrak{sl}_n]\to \cY_{\nu_1}^{\lambda_1}[\mathfrak{sl}_n]\otimes \cY_{\nu_2}^{\lambda_2}[\mathfrak{sl}_n]$. We are only concerned with the balanced case, where the shift $\nu$ vanishes. The existence of this map will play an important role for us.

Finally, in Section \ref{sec:traces} we leverage the morphism from the Yangian to argue that correlators and the corresponding twisted trace can be written in terms of transfer matrices of the inhomogeneous $\mathfrak{sl}_n$ spin chain. For the $\cA_{N;n}\equiv \cA^{(0)}_{N;n}$ algebra in the $n=2$ case (that is $N$ D3 branes intersecting 2 D5 or NS5 branes), we fully solve the problem by proposing an explicit formula for the correlators. We achieve this by combining the coproduct property \cite{WeekesThesis} with the localization results of \cite{Dedushenko:2017avn,Dedushenko:2018icp}. First we argue for \eqref{answer_3d}, which gives a remarkably explicit answer for the correlators in 3d $\cN=4$ SQCD with the gauge group $U(N)$ and $2N$ fundamental flavors. It is written in terms Q-operators of the XXX spin chain. We then derive the equation \eqref{answer_4d}, which computes correlators of the interface operators in terms of transfer matrices. In both cases, we are concerned with the correlator $u_1^N\dots u_L^N\langle T[u_1]^{a_1}_{b_1}\dots T[u_L]^{a_L}_{a_L}\rangle$, where $T[u]^a_b$ is the generating function of the $\cY[\mathfrak{sl}_2]$ generators in the former and $\cY[\mathfrak{gl}_2]$ in the latter case. The correlator is computed as a matrix element
\begin{equation}
u_1^N\dots u_L^N\langle T[u_1]^{a_1}_{b_1}\dots T[u_L]^{a_L}_{a_L}\rangle=\big\langle a_1,\dots,a_L;L\big|M_L\big|b_1,\dots,b_L;L\big\rangle
\end{equation}
of a certain $2^L\times 2^L$ matrix $M_L$ built with the help of a spin chain data. Here $|a_1,\dots,a_L;L\rangle$ denotes the standard spin chain basis, with $|1,\dots,1;L\rangle$ meaning $|\uparrow,\dots,\uparrow\rangle$, $|1,\dots,1,2;L\rangle$ meaning $|\uparrow,\dots,\uparrow\downarrow\rangle$, etc. In the case of 3d SQCD, the masses are fixed, and the matrix is
\begin{equation}
M_L=\sum_{\sigma\in\substack{\frac{S_{2N}}{S_{N}\times S_N}}} \frac{i^{-N^2} e^{i\pi \zeta\sum_{j=1}^{2N} \mu_j} }{(2\sinh\pi\zeta)^N\prod_{a=1}^N \prod_{k=N+1}^{2N}2\sinh \pi(\mu_{\sigma(a)}-\mu_{\sigma(k)})}\prod_{a=1}^N \mathbf{Q}_+(\vec{u}-\mu_{\sigma(a)}\vec{e})\mathbf{Q}_-(\vec{u}-\mu_{\sigma(a+N)}\vec{e}),
\end{equation}
where $\mu_1,\dots,\mu_{2N}$ are dimensionless masses $\mu_j=\ell m_j$ (here $\ell$ is the sphere radius), $\zeta$ is the FI parameter that also defines twisted-periodic boundary conditions for the XXX spin chain, $\vec{u}=(u_1,u_2,\dots,u_L)$, $\vec{e}=(1,1,\dots,1)$, and $\mathbf{Q}_\pm$ are Baxter Q-operators of the inhomogeneous length-$L$ XXX spin chain with the inhomogeneities $\vec{u}$. The sum runs over the massive vacua. In the interface case, the masses are integrated over and promoted to central elements of $\cY[\mathfrak{gl}_2]$ (see also \cite{Finkelberg_2019}), whose trace is encoded in the matrix $M_L$ that takes a different form:
\begin{align} \label{answer}
M_L=\frac1{N!}\int_{\R^N\times\R^N}&[\dd \mu^L][\dd \mu^R] e^{-\frac{i\pi}{\tau}\tr(\mu^L)^2 - \frac{i\pi}{\tau}\tr(\mu^R)^2+2\pi i\zeta\sum_{a=1}^N \bar{\mu}_a}\Delta(\mu^L)\Delta(\mu^R)\cr
&\times\prod_{j=1}^N \frac{\mathbb{T}^+_{-\frac12 +i(\mu^L_j-\bar\mu_j)}(\vec{u}-\bar\mu_j \vec{e})-\mathbb{T}^+_{-\frac12 +i(\mu^R_j-\bar{\mu}_j)}(\vec{u}-\bar{\mu}_j\vec{e})}{2i\sinh\pi(\mu^L_j-\mu^R_j)},
\end{align}
where $\bar\mu_a=\frac12(\mu^L_a +\mu^R_a)$, and the integral implements coupling to the bulk on the two sides of the interface. Here $\mathbb{T}^+_j(\vec{u})$ stands for the transfer matrix (for the $\mathfrak{sl}_2$ Verma module of highest weight $j$) of the same spin chain. The transfer matrices of this kind are expressed through the Baxter Q-operators, and both are easily computable using the results of \cite{Bazhanov:2010ts} (see \cite{Frassek:2018try} for generalizations), making the above answers very explicit. Thus we fully solve the purely 3d problem, while the interface answer \eqref{answer} still involves a complicated integral. We should note that for the Verma module transfer matrices to make sense, we must keep the twist parameter generic until the end, which in the NS5 description is the interface FI term $\zeta$. The final answer must admit the $\zeta\to 0$ limit, even if individual transfer matrices might diverge in this limit. Quite conveniently, the difference of transfer matrices that enters the product in \eqref{answer} is precisely the combination that is known to admit a finite $\zeta\to0$ limit \cite{Bazhanov:2010ts}.

Appendices contain some technical material for the Sections \ref{sec:Higgs_alg} and \ref{sec:traces}.

\subsection{Outlook}

A straightforward generalization of \eqref{answer} would be to extend it for $\cA_{N;n}$ with $n>2$. The answer would express correlators of operators from $\cA_{N;n}$ in terms of a particular linear combination of products of transfer matrices for the $\mathfrak{sl}_n$ spin chain. In this case the results of \cite{Bazhanov:2010jq} are of use. It would be interesting to work out such a generalization. The case with different numbers of D3 branes terminating on fivebranes, which is expected to be controlled by shifted Yangians, is another interesting generalization.

The formula \eqref{answer} serves as a natural starting point for the large-N analysis of correlators. As it stands, it suggests a neat interpretation 
for the generating functions $T[u]^a_b$ as dual to ``giant open strings'' attached to the D5 branes, with endpoints describing a 
macroscopic Wilson line in the five-brane theory. In this protected sector, the $T[u]^a_b$ vevs measure some overall holonomy along the Wilson line, 
induced by the backreaction of the D3 branes in the D5 brane worldvolume. The backreaction is represented by a collection of $N$ 't Hooft lines. 
We plan to address the large $N$ limit elsewhere.

\subsection*{Acknowledgments}
We thank K.Costello and N.Nekrasov for useful discussions and A.Weekes for explaining to us several important points about quantized Coulomb branch algebras. This research was supported in part by a grant from the Krembil Foundation. D.G. is supported by the NSERC Discovery Grant program and by the Perimeter Institute for Theoretical Physics. Research at Perimeter Institute is supported in part by the Government of Canada through the Department of Innovation, Science and Economic Development Canada and by the Province of Ontario through the Ministry of Colleges and Universities.

\section{Higgs or D5 presentation}\label{sec:Higgs_alg}
In this section we explore algebras describing 1d, or TQM, sectors on interfaces engineered by $n$ D5 branes. They intersect $N$ D3 branes, and there may be extra $ns$ D3 branes that terminate on D5 branes from the right. When $s=0$, so there are only intersecting branes, the algebra is denoted $\cA_{N;n}$. When $s>0$, we assume that exactly $s$ D3 branes terminate on each D5 brane, and the corresponding algebra is denoted $\cA^{(s)}_{N;n}$. The cases $s=0$, $s=1$, and $s>1$ are treated separately: only the first one involves extra hypermultiplets living at the interface, while the second is built from regular boundary conditions, and the latter case involves the Nahm pole $\varrho=[1^N, s^n]$.

\subsection{$N$ D3 branes crossing $n$ D5 branes}\label{sec:N_cross_n}
The 4d QFT description of the interface is straightforward: $n$ fundamental 3d hypermultiplets 
coupled to the 4d $U(N)$ gauge fields. The 3d hypermultiplets also couple in the obvious way to 
the IR free 6d $U(n)$ SYM living on the D5 branes worldvolume. 

At the level of 4d SYM with the interface, the sphere background (or the $\Omega$ deformation) reduces the 3d hypermultiplets to a 1d quantum mechanical system, 
a collection of $nN$ Weyl algebras. This is coupled to a perturbative 2d $\mathfrak{gl}(N)$ YM theory arising from the $\Omega$ deformation
of 4d SYM, in a manner discovered in \cite{Pestun:2009nn} and described in the present context in \cite{DG1} (see also \cite{Wang:2020seq}). The result is a local operator algebra ${\cal A}_{N;n}$
obtained as the quantum Hamiltonian reduction of the collection of Weyl algebras combined with 
two copies of the universal enveloping algebra $U[\mathfrak{gl}_n]$. That is, we start with
\begin{equation}
U(\mathfrak{gl}_N) \otimes {\rm Weyl}^{Nn} \otimes U(\mathfrak{gl}_N),
\end{equation}
where the two copies of $U(\mathfrak{gl}_N)$ are generated by $(B_\pm)^\alpha_\beta$, and the $Nn$ copies of Weyl algebra are generated by $X^a_\alpha$, $Y_a^\alpha$, where the lowercase Greer letters denote $\mathfrak{gl}_N$ indices (in most formulas, we will suppress them, unless needed). The commutators obeyed by the generators are
\begin{align}
[(B_\pm)^\alpha_\beta, (B_\pm)^\gamma_\delta]&=\hbar \delta^\gamma_\beta (B_\pm)^\alpha_\delta - \hbar \delta^\alpha_\delta (B_\pm)^\gamma_\beta,\cr
[X^a_\alpha, Y_b^\beta] &= \hbar \delta^a_b \delta_\alpha^\beta,
\end{align}
where we conveniently introduced an explicit quantization parameter $\hbar$ of weight $2$. We can give weight $2$ to the generators $B_+$ and $B_-$, and weight $1$ to the generators $X^a$ and $Y_a$. We then perform the quantum Hamiltonian reduction: first take the quotient by the left ideal generated by the F-term relation
\begin{equation}
\mu^\alpha_\beta \equiv (B_+)^\alpha_\beta + (B_-)^\alpha_\beta + Y^\alpha_a X^a_\beta + \hbar N \delta^\alpha_\beta =0,
\end{equation}
and then restrict to $\mathfrak{gl}_N$ invariants. The FI parameter can be absorbed into the diagonal components of $B_\pm$, but we fixed it to a convenient value. In this way, we construct the algebra $\cA_{N;n}$.

The algebra $\cA_{N;n}$ has a large commutative sub-algebra generated by operators of the form $\Tr B_+^k$ and operators of the form 
$\Tr B_-^k$. This subalgebra is actually central: we can use the F-term relations to eliminate $B_-$ from a generic operator
to show that it commutes with all $\Tr B_-^k$, and similarly for $B_+$. We will denote these central generators as
\begin{equation}
b_{+,k} = \Tr B_+^k \qquad \qquad b_{-,k} = \Tr B_-^k.
\end{equation}
They are the bulk operators in the $U(N)$ 4d SYM on the left and on the right of the D5 branes: just like in the case of boundaries \cite{DG1}, the center of the interface algebra is generated by the bulk operators.

Now we would like to identify the morphism from $\cY[\mathfrak{gl}_n]$ to $\cA_{N;n}$. The rough form of the morphism is really determined by the physics: 
the generators $B_\pm$ in the two copies of $U[\mathfrak{gl}_N]$ arise from the 4d SYM scalar fields which control the transverse position of the D3 branes in the holomorphic plane of the 4d CS theory. Thus the $n$-th derivative of the connection should couple to 
generators of the form $X B_\pm^n Y$, with $X$ and $Y$ being the generators of the Weyl algebra. We verify in Appendix \ref{app:yang} that these are indeed Yangian generators. 

More precisely, the $\cY[\mathfrak{gl}_n]$ Yangian has an RTT presentation, which employs the monodromy matrix $T[u]$ 
associated to the fundamental representation. Written as a commutator, it takes the form 
\begin{equation}
 [T[u]^a_b, T[v]^c_d] =  \hbar \frac{T[v]^c_b  T[u]^a_d-T[u]^c_b T[v]^a_d}{u-v} ,
\end{equation}
where 
\begin{equation}
T[u] \equiv 1 + \frac{t^{[1]}}{u} + \frac{t^{[2]}}{u^2} + \cdots
\end{equation}
is a formal generating function for the Yangian generators. Notice that the RTT relations for the $\cY[\mathfrak{gl}_n]$ Yangian do not really constrain the overall scale of $T[u]$, which can be redefined by multiplication by any scalar formal power series in $1+u^{-1}\C[[u^{-1}]]$, possibly with coefficients living in the center of the Yangian. 

It is also useful to define an inverse formal series by 
\begin{equation}
T[u]^a_b \bar T[u]^b_c  = \delta^a_c,
\end{equation}
which gives the monodromy matrix associated to the anti-fundamental representation. 

Then we find identifications:
\begin{equation}
T[u]^a_b = \delta^a_b - X^a \frac{1}{u-B_+} Y_b,  \qquad \qquad \bar T[u]^a_b = \delta^a_b + X^a \frac{1}{u+B_-} Y_b,
\end{equation}
mapping the Yangian generators to ${\cal A}_{N;n}$. In these expressions 
$B_\pm$ are treated as $N \times N$ matrices of operators, and contraction of $\mathfrak{gl}_N$ indices is implied. More explicitly, 
 \begin{equation}
t^{[n]}{}^a_b = -X_{\alpha_1}^a (B_+)^{\alpha_1}_{\alpha_2} \cdots (B_+)^{\alpha_{n-1}}_{\alpha_n} Y^{\alpha_n}_b. 
\end{equation}
That $T[u]$ and $\bar{T}[u]$ indeed obey the Yangian relations is shown in the Appendix \ref{app:yang}.

The images of the Yangian generators $t^{[n]}{}^a_b$ generate ${\cal A}_{N;n}$, but satisfy extra relations, which truncate the Yangian to a finitely-generated algebra $\cA_{N;n}$.  
Indeed, a nice property of $U[\mathfrak{gl}_N]$ is the existence of a degree $N$ characteristic polynomial $P_+(u)$, 
whose coefficients are elements in the center of $U[\mathfrak{gl}_N]$, such that $P_+(B_+)=0$.\footnote{It is given by the Capelli determinant of $u-B_+$ \cite{Capelli}, see \cite{Molev} for details on the relevant algebraic techniques.} That means $P_+(u)\frac{1}{u-B_+}=(P_+(u)-P_+(B_+))\frac{1}{u-B_+}$ is a degree $N-1$ polynomial in $u$, and thus $P_+(u) T[u]^a_b$ is a degree $N$ 
polynomial in $u$ starting with $u^N  \delta^a_b$. We can similarly define a $P_-(u)$ such that $P_-(-B_-)=0$,
which controls the denominator of $\bar T[u]$ and the corresponding Yangian truncation.

The center of the $\cY[\mathfrak{gl}_n]$ Yangian is given by the coefficients of the quantum determinant of $T[u]$. We have not directly proven in this presentation, but we have tested the statement that the quantum determinant of $T[u]^a_b$ should be $\frac{P_-(u-\frac{n-1}{2}\hbar)}{P_+(u+ \frac{n-1}{2} \hbar)}$ and the quantum determinant of $\bar T[u]^a_b$ should be $\frac{P_+(u+\frac{n-1}{2}\hbar)}{P_-(u- \frac{n-1}{2} \hbar)}$, see Appendix \ref{app:yang}. That means the central elements of the Yangian are determined 
as a function of the central elements of the two copies of $U[\mathfrak{gl}_N]$, i.e. the bulk operators.

\subsubsection{Fermionic fundamental fields and Wilson lines} \label{wilson}
There is an alternative embedding of the D5 branes which is compatible with the $\Omega$ deformation/sphere background: they can be taken to share 
a single space-time direction with the D3 branes. The $35$ open strings give rise to a collection of $Nm$ complex fermions, effectively engineering 
a BPS Wilson line in the $\Lambda \mathbb{C}^{mN}$ representation of $U(N) \times U(m)$ \cite{Yamaguchi:2006tq}. 

After localization, the effect of the complex fermions is completely analogous to that of the $X^a,Y_b$ fields above, except that 
they are Grassmann-odd and define a Clifford algebra rather than a Weyl algebra. Indeed, calculations in \cite{Ishtiaque:2018str} were done for the case of fermions. We can also consider a combination of a composite Wilson line and an interface,
so that we have $n$ bosonic and $m$ fermionic $X^a,Y_b$ fields. 

All the algebraic calculations above work in exactly the same manner, giving a map from the $\cY[\mathfrak{gl}_{n|m}]$ Yangian
to the resulting algebra ${\cal A}_{N;n|m}$ of local operators. We leave the details to future work. 

\subsection{$n$ D5 branes between $N$ and $N+n$ D3 branes}
This interface is simpler: the gauge group is reduced from $U(N+n)$ to $U(N)$ at the interface, 
with the commutant $U(n)$ being the interface global symmetry.

The 1d algebra is the $\mathfrak{gl}_N$ quantum Hamiltonian reduction of  $U[\mathfrak{gl}_{N+n}]\times U[\mathfrak{gl}_N]$. The moment map constraint simply kills the 
$U[\mathfrak{gl}_N]$ factor, leaving the $\mathfrak{gl}_N$-invariant part of $U[\mathfrak{gl}_{N+n}]$. We denote the resulting algebra as ${\cal A}^{(1)}_{N;n}$.
 
One can construct several homomorphisms from the $\cY[\mathfrak{gl}_n]$ Yangian to ${\cal A}^{(1)}_{N;n}$ following \cite[Section 2.13]{Molev}.  One is a well-known  map from $\cY[\mathfrak{gl}_n]$ to ${\cal A}^{(1)}_{N;n}$, denoted in \cite[Section 2.13]{Molev} as
\begin{equation}
\psi_{N+n}: \cY[\mathfrak{gl}_n] \to \cA^{(1)}_{N;n}
\end{equation} 
To construct it, one takes the matrix $1+ \frac{B_+}{u}$, with $B_+\in\mathfrak{gl}_{N+n}$ (which is an image of the $\cY[\mathfrak{gl}_{N+n}]$ Yangian $T$) and computes quantum determinants 
of submatrices of size $N+1$ with indices $(a,n+1, \cdots n+N)$ and $(b,n+1, \cdots n+N)$, where $a,b=1..k$. The result is the image of $T[u]^a_b$.

Other homomorphisms can also be constructed based on the evaluation map from $\cY[\mathfrak{gl}_{N+n}]$ to $U(\mathfrak{gl}_{N+n})$ given by $1+\frac{B_+}{u}$. The simplest thing we can do is restrict this map to a $\cY[\mathfrak{gl}_n]$ subalgebra, which gives the evaluation map again, now to $U(\mathfrak{gl}_n)\subset U(\mathfrak{gl}_{N+n})^{\mathfrak{gl}_N}=\cA^{(1)}_{N;n}$.   However, there exists a more interesting homomorphism from $\cY[\mathfrak{gl}_n]$, which was denoted $\varphi_{N+n}$ in \cite[Section 2.13]{Molev}, 
\begin{equation}
\varphi_{N+n}: \cY[\mathfrak{gl}_n] \to \cA^{(1)}_{N;n},
\end{equation}
defined by restricting the following map,
\begin{equation}
\label{secondHom}
\bar T[u] \mapsto \left(1 + \frac{B_+}{u-a} \right)^{-1},
\end{equation}
to the $\cY[\mathfrak{gl}_n]$ subalgebra. Here $a$ is an arbitrary number, but to match \cite{Molev}, one should take $a=N+n$ (and replace 
$\bar T[u]$ with $T[-u]$). It is easy to see using the Yangian relations that this restriction is manifestly $\mathfrak{gl}_N$-invariant, thus indeed giving a homomorphism from $\cY[\mathfrak{gl}_n]$ to $\cA^{(1)}_{N;n}$.

Based on explicit tests at small $N$ and some commutative limits, we expect the two maps $\psi_n$ and $\varphi_n$ to actually coincide. 
 
 \subsection{$n$ D5 branes between $N$ and $N+n s$ D3 branes}
This setup is a bit more complicated to analyze: the gauge group is reduced from $U(N+n s)$ to $U(N)$ at the interface,
but the commutant $U(n s)$ is further broken to $U(n)$ by a Nahm pole consisting of $n$ blocks of size $s$. 

At the level of the algebra, that means that we need the $\mathfrak{gl}_N$-invariant part of a finite W-algebra $\cW(\mathfrak{gl}_{N+ns},e)$ built from the quantum Drinfeld-Sokolov (DS) reduction of 
$U[\mathfrak{gl}_{N+n s}]$ involving a nilpotent generator $e$ with $n$ Jordan blocks of size $s$. We denote the resulting algebra as ${\cal A}^{(s)}_{N;n}$.
It seems to be possible to also describe it as the quantum DS reduction of ${\cal A}^{(1)}_{N;n s}$. \footnote{It is natural to wonder if the quantum DS reduction of
the $\cY[\mathfrak{gl}_{n s}]$ Yangian involving a nilpotent generator with $n$ Jordan blocks of size $s$ would admit a map from (or coincide with) the $\cY[\mathfrak{gl}_{n}]$ Yangian.}

It turns out that the description of $\cA^{(s)}_{N;n}$ as the subalgebra of invariants,
\begin{equation}
\cA^{(s)}_{N;n}=\cW(\mathfrak{gl}_{N+ns},e)^{\mathfrak{gl}_N},
\end{equation}
can be made quite explicit. The Jordan type of $e$ is determined by
\begin{equation}
\varrho=[\underbrace{1,\dots,1}_N,\underbrace{s,\dots, s}_{n}].
\end{equation}
We characterize the algebra $\cW(\mathfrak{gl}_{N+ns},e)$ using its relation to shifted Yangians \cite{BRUNDAN2006136}. As a first step, choose a good grading compatible with $e$. Good gradings on $\mathfrak{gl}$ algebras are classified in terms of pyramids \cite{elashvili2003good}, and a convenient choice of pyramid for our $\varrho$ is
\begin{center}
\ytableausetup{mathmode, boxsize=2em}
\begin{ytableau}
	\substack{1} \cr
	.\cr
	.\cr
	\substack{N}\cr
	\substack{N+1} & . & . & . & \substack{N+s}\cr
	.   & . & . & . & . \cr
	.   & . & . & . & . \cr
	.   & . & . & . & \substack{N+ns} \cr
\end{ytableau}
\end{center}
The corresponding shift matrix is given by
\begin{equation}
\sigma = \left(\begin{matrix}
0 & 0 & \dots & 0 & s-1 & \dots & s-1\\
0 & 0 & \dots & 0 & s-1 & \dots & s-1\\
. & . & \dots & . & .   & \dots & . \\
0 & 0 & \dots & 0 & s-1 & \dots & s-1\\
0 & 0 & \dots & 0 & 0   & \dots & 0 \\
0 & 0 & \dots & 0 & 0   & \dots & 0 \\
. & . & \dots & . & .   & \dots & . \\
0 & 0 & \dots & 0 & 0   & \dots & 0 \\
\end{matrix} \right).
\end{equation}
It is an $(N+n)\times(N+n)$ matrix, consisting of three zero blocks of sizes $N\times N$, $n\times N$ and $n\times n$, with the only non-zero block of size $N\times n$ with $s-1$ in every entry. It follows from the results of \cite{BRUNDAN2006136} that the most convenient description of the corresponding shifted and truncated Yangian $Y_{N+n,s}(\sigma)$ is via the parabolic presentation of shape
\begin{equation}
\label{shape}
\nu=(N,n).
\end{equation}
Using \cite[Corollary 6.3]{BRUNDAN2006136}, we identify the generators in their notations as
\begin{equation}
\label{gen_N_ns}
\{D^{(1)}_{1; i,j}\}_{1\leq i,j \leq N},\quad \{D^{(r)}_{2; a,b}\}_{1\leq a,b\leq n, 1\leq r \leq s},\quad \{E^{(s)}_{1;i,a}\}_{1\leq i\leq N, 1\leq a\leq n},\quad \{F^{(1)}_{1; b,j}\}_{1\leq b\leq n, 1\leq j\leq N},
\end{equation}
which generate $\cW(\mathfrak{gl}_{N+ns},e)$, and ordered polynomials in which form its PBW basis. The relations obeyed by these generators are written in \cite[(3.3)--(3.14)]{BRUNDAN2006136}. The first set, $\{D^{(1)}_{1; i,j}\}_{1\leq i,j \leq N}$, generates the $U(\mathfrak{gl}_N)$ subalgebra, and we are supposed to take invariants with respect to it, i.e. the next step is to compute
\begin{equation}
\cA^{(s)}_{N;n}=\cW(\mathfrak{gl}_{N+ns},e)^{\mathfrak{gl}_N}.
\end{equation}
The second set in \eqref{gen_N_ns} commutes with the first set, while the remaining two transform as the fundamental and the anti-fundamental of $\mathfrak{gl}_n$ respectively. This allows to identify the $\mathfrak{gl}_N$-invariant generators as follows,
\begin{equation}
\label{glN_inv_gen}
\{D^{(r)}_{2; a,b}\}_{1\leq a,b\leq n, 1\leq r \leq s},\quad \tr (D^{(1)}_1)^m,\quad \{F^{(1)}_{1;a} (D^{(1)}_{1})^m E^{(s)}_{1;b}\}_{1\leq a,b\leq n,\ m\in \Z_{\geq 0}},
\end{equation}
where in the last set of generators we suppressed summation over the $\mathfrak{gl}_N$ indices $i,j=1..N$.

Notice that the generators from the first set in \eqref{glN_inv_gen} obey the Yangian $\cY[\mathfrak{gl}_n]$ relations:
\begin{equation}
\label{subYang}
[D^{(r)}_{2;a,b}, D^{(p)}_{2;c,d}]=\sum_{t=0}^{\min(r,p)-1} \left( D^{(r+p-1-t)}_{2;a,d}D^{(t)}_{2;c,b} - D^{(t)}_{2;a,d}D^{(r+p-1-t)}_{2;c,b} \right).
\end{equation}
Of course, the elements $D^{(r)}_{2;a,b}$ with $r>s$ are not independent generators, for they are not on the list \eqref{gen_N_ns}. This is the effect of truncation, -- the algebra $Y_{N+n,s}(\sigma)$ is defined as a quotient of the shifted Yangian over a two-sided ideal generated by $\{D^{(r)}_{1;i,j}\}_{1\leq i,j\leq N, r>1}$. As a result, elements $D^{(r)}_{2;a,b}$ with $r>s$ can be expressed, using the relations \cite[(3.3)--(3.14)]{BRUNDAN2006136}, as polynomials in generators listed in \eqref{gen_N_ns}. This is a straightforward somewhat technical exercise, which we do not present here for the sake of brevity. The consistency of truncation implies that the relations \eqref{subYang} are obeyed by all $D^{(r)}_{2;a,b}$, even if some of them are not independent. In particular, this implies a homomorphism:
\begin{align}
\cY[\mathfrak{gl}_n] &\to \cA^{(s)}_{N;n},\cr
t^{[r]}{}^a_b &\mapsto D^{(r)}_{2;a,b}.
\end{align}
The simplicity with which we are able to define this homomorphism is due to the optimal choice of the parabolic presentation shape \eqref{shape}, thus all the technicalities are hidden in \cite{BRUNDAN2006136}.

We now shift gear to the Coulomb branch description of the algebra.
 
\section{The Coulomb or NS5 perspective}\label{sec:Coulomb}
\begin{figure}[t]
	\label{fig:quiver}
	\centering
	\includegraphics[scale=0.6]{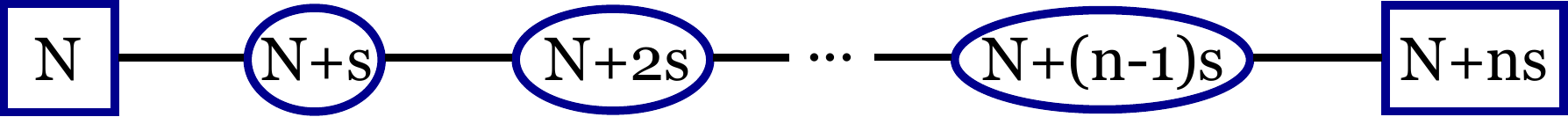}
	\caption{The balanced quiver that provides the Coulomb branch description of $\cA^{(s)}_{N;n}$.}
\end{figure}

S-duality gives an alternative presentation of ${\cal A}^{(s)}_{N;n}$: the quantum Coulomb branch algebra of an NS5 interface containing n NS5 branes, with D3 branes crossing and terminating on them just like in the D5 case. It is describe by a balanced linear quiver of $(n-1)$ unitary gauge groups, 
with $N$ flavors at one end and $N+n s$ at the other, see Figure \ref{fig:quiver}. The ranks at the $i$-th gauge node are $N+ i s$, and the flavor nodes are coupled to the bulk on the two sides of the interface, so the masses for the flavors are promoted to central elements. Unlike in the D5 frame, the description is uniform in $s$: there is no need to consider $s=0$, $s=1$, and $s>1$ separately. The algebra of the linear quiver in Figure \ref{fig:quiver} is a central quotient of $\cA^{(s)}_{N;n}$, and we sometimes denote it as $\cA^{(s)}_{N;n}[m]$, where $m$ stands for masses that parameterize the quotient. It is given by the truncated Yangian $\cY_0^\lambda[\mathfrak{sl}_n]$, where $\lambda$ is the truncation parameter.

The correspondence between Coulomb branches of balanced ADE quivers and truncations of the corresponding ADE Yangian 
is well understood \cite{WeekesThesis}. The specific truncation depends on the number and location of the flavours on the quiver. The balanced condition essentially means that the 
vector $(N_i)$ built from the number of flavours at each node equals the ADE Cartan matrix acting on the vector $(M_i)$ built from the gauge ranks. 

The map from the Yangian means that the corresponding quantized Coulomb branch algebra $\cY_0^\lambda[\mathfrak{sl}_n]$ can always be employed to define a line defect in 4d ADE CS theory. 
The line defect is recognized in \cite{CGY} as a collection of 't Hooft operators in one-to-one correspondence with the flavours of the quiver: 
the (minuscule) charge is controlled by the position of the flavour in the ADE quiver and the locations/spectral parameter is the corresponding mass parameter.  

In the general ADE case, the masses are thus not identified with central elements in the Yangian: the ADE Yangian has no center (in fact for any simply $\mathfrak{g}$, the Yangian $\cY[\mathfrak{g}]$ has no center \cite{Wendlandt_2018}). 
Instead, they control the specific truncation of the Yangian. In the case at hand associated to the D3-D5 interfaces (and their duals), though, 
because the brane interpretation involves a $U(n)$ 6d gauge theory, we expect the map from the $\cY[\mathfrak{sl}_{n}]$ Yangian to the quantum Coulomb branch of the linear quiver to admit some natural lift to a map from the $\cY[\mathfrak{gl}_{n}]$ Yangian to the quantum Coulomb branch of the D3-NS5 {\it interface} obtained by coupling the 3d linear quiver theory to the 
D3 brane SYM, i.e. by promoting masses to central generators. Indeed, see \cite[Theorem 2.34]{frassek2020lax} on such a lift.

Recall that the RTT presentation of $\cY[\mathfrak{sl}_{n}]$ is obtained from the RTT presentation of $\cY[\mathfrak{gl}_{n}]$ by imposing a quantum determinant 
constraint ${\rm qdet}\, T[u] = 1$ \cite{GuideToQu}. One can adjust the scale of $T[u]$ by some functions of the masses to get a more general normalization of the quantum determinant,
promoting the map from $\cY[\mathfrak{sl}_{n}]$ to the quantum Coulomb branch to a map  from $\cY[\mathfrak{gl}_{n}]$ to the  quantum Coulomb branch extended by the masses. 
Based on the $s=0$ example and on co-product considerations we discuss below we expect the natural normalization of $T[u]$ to be such that 
$Q_0(u) T[u]$ and $Q_n(u) \bar T[u]$ are polynomials in $u$, where 
\begin{equation}
Q_0(z)=\prod_{a=1}^N(z-m^L_a),\quad Q_n(z)=\prod_{a=1}^{N+ns} (z-m^R_a).
\end{equation}
We will avoid using this expectation in concrete calculations, though.

Quantum Coulomb branch algebras for the linear quivers can be explicitly described \cite{Bullimore:2015lsa, Braverman:2016pwk}, in general giving truncations of shifted Yangians, with shifts vanishing for our balanced quivers. The traces can be written using the methods of \cite{Dedushenko:2017avn,Dedushenko:2018icp}. The basic building blocks are the shift operators:
\begin{align}
\label{shift_op}
u_{i,a}^+ &= -\frac1\ell \frac{\prod_{b=1}^{M_{i-1}} (\frac12 + i\ell(\phi_{i,a}-\phi_{i-1,b}))}{\prod_{b\neq a}i\ell(\phi_{i,a}-\phi_{i,b})} e^{-\frac{i}{2}\partial_{\sigma_{i,a}}-\partial_{B_{i,a}}}= \frac{-i}{(i\ell)^{M_{i}-M_{i-1}}}\frac{\prod_{b=1}^{M_{i-1}} (\phi_{i,a}-\phi_{i-1,b}-\frac{\epsilon}{2})}{\prod_{b\neq a}(\phi_{i,a}-\phi_{i,b})} e^{-\epsilon\partial_{\phi_{i,a}}},\cr
u_{i,a}^- &= \frac1\ell \frac{\prod_{b=1}^{M_{i+1}} (\frac12 + i\ell(\phi_{i+1,b}-\phi_{i,a}))}{\prod_{b\neq a}i\ell(\phi_{i,b}-\phi_{i,a})} e^{\frac{i}{2}\partial_{\sigma_{i,a}}+\partial_{B_{i,a}}}=\frac{-i}{(-i\ell)^{M_{i}-M_{i+1}}} \frac{\prod_{b=1}^{M_{i+1}} (\phi_{i,a}-\phi_{i+1,b}+\frac{\epsilon}2)}{\prod_{b\neq a}(\phi_{i,a}-\phi_{i,b})} e^{\epsilon\partial_{\phi_{i,a}}},\cr
\phi_{i,a} &=\frac1{\ell} (\sigma_{i,a} + \frac{i}{2} B_{i,a}),
\end{align}
that act on functions of $\sigma_{i,a}\in\R$ and $B_{i,a}\in\Z$, which are coordinates on the Cartan subalgebra of the gauge group $U(M_1)\times\dots\times U(M_{n-1})$ and its cocharacter lattice respectively. We also introduced the complex combinations $\phi_{i,a}$ of these coordinates. The notation $M_j = N + js$ was used for the ranks of gauge groups, and $\ell$ is the sphere radius (in the $S^3$ background approach), which then translates to $\epsilon=\frac{i}{\ell}$. We also identified the flavor nodes as the $0$th and $n$th ``gauge nodes'', so $\phi_{0,a}=m^L_a$ and $\phi_{n,a}=m^R_a$ are masses. The shift operators $u^\pm_{i,a}$ in \eqref{shift_op} correspond to abelianized minuscule monopoles, and together with $\phi_{i,a}$ they generate the abelianized Coulomb branch algebra $\cA^{\rm ab}_C[\cT]$ of our 3d quiver theory $\cT$, whose Weyl-invariant subalgebra is the true Coulomb branch algebra of the linear quiver, $\cA_C[\cT]=(\cA_C^{\rm ab}[\cT])^\cW$.  One can then define the generating functions of Weyl-invariant generators following \cite{Bullimore:2015lsa}:
\begin{align}
\label{Coul_Yang}
Q_i(z)&=\prod_{a=1}^{M_i} (z-\phi_{i,a}),\cr
H_i(z)&=\frac{Q_{i-1}(z+\frac{\epsilon}{2})Q_{i+1}(z+\frac{\epsilon}{2})}{Q_i(z)Q_i(z+\epsilon)},\quad U_i^\pm(z)=\sum_{a=1}^{M_i} u^\pm_{i,a}\prod_{b\neq a}(z-\phi_{i,b}),\quad i=1\dots n-1,\cr
E_i(z) &= \frac1{Q_i(z)}U^-_i(z)=\sum_{a=1}^{M_i} \frac1{z-\phi_{i,a}}u^-_{i,a},\cr
F_i(z)&=U^+_i(z) \frac1{Q_i(z)}=\sum_{a=1}^{M_i} \frac1{z-\phi_{i,a}+\epsilon}u^+_{i,a}.
\end{align}
It is straightforward to check that (assuming the quiver is balanced)
\begin{align}
[\phi_{i,a},\phi_{j,b}]&=0,\cr
[\phi_{j,b},u^\pm_{i,a}]&=\pm\epsilon\delta_{i,j}\delta_{a,b}u^\pm_{i,a},\cr
[u^+_{i,a}, u^-_{j,b}]&\propto \delta_{i,j}\delta_{a,b},\cr
u^+_{i,a}u^-_{i,a}&=-\frac{Q_{i-1}(\phi_{i,a}-\frac{\epsilon}2)Q_{i+1}(\phi_{i,a}-\frac{\epsilon}2)}{\prod_{b\neq a}(\phi_{i,a}-\phi_{i,b})(\phi_{i,a}-\phi_{i,b}-\epsilon)},\cr
u^-_{i,a}u^+_{i,a}&=-\frac{Q_{i-1}(\phi_{i,a}+\frac{\epsilon}2)Q_{i+1}(\phi_{i,a}+\frac{\epsilon}2)}{\prod_{b\neq a}(\phi_{i,a}-\phi_{i,b})(\phi_{i,a}-\phi_{i,b}+\epsilon)}.
\end{align}
Again, $Q_0$ and $Q_n$ correspond to the flavor nodes, encoding the corresponding masses:
\begin{equation}
Q_0(z)=\prod_{a=1}^N(z-m^L_a),\quad Q_n(z)=\prod_{a=1}^{N+ns} (z-m^R_a).
\end{equation}
The above equations, with shift operators defined in \eqref{shift_op}, exactly match those of the well-known GKLO representation \cite{Gerasimov_2005}. Therefore the generating series $H_i(z)$, $E_i(z)$, $F_i(z)$ form a representation of the Yangian $\cY[\mathfrak{sl}_n]$. Indeed, \cite{Bullimore:2015lsa} check that
\begin{align}
[H_i(z), E_j(w)]&= -\frac{\epsilon}{2} \kappa_{ij} \frac{[H_i(z), E_j(z)-E_j(w)]_+}{z-w},\cr
[H_i(z), F_j(w)]&=-\frac{\epsilon}{2} \kappa_{ij} \frac{[H_i(z), F_j(z)-F_j(w)]_+}{z-w},\cr
[E_i(z), F_j(w)]&=-\epsilon \delta_{ij} \frac{H_i(z) - H_i(w)}{z-w},
\end{align}
where $\kappa_{ij}$ is the Cartan matrix of $\mathfrak{sl}_n$. More precisely, this is a representation of a specialization of $Y[\mathfrak{gl}_n]$, where the quotient sets elements of the center to certain numeric values determined by the masses $(m^L, m^R)$. It is straighforward to adjust the normalization of $T[u]$ as indicated above, so that it has denominator $Q_0(u)$ 
while  $\bar T[u]$ has denominator $Q_n(u)$

Now we can finally construct the trace. For that we define the ``vacuum'' wave function (or ``empty hemisphere'' partition function) $\Psi_0(\sigma,B)$, which is a function on $\mathfrak{t}\times \Lambda^\vee$ \cite{Dedushenko:2017avn,Dedushenko:2018icp}:\footnote{$\Lambda^\vee$ is the cocharacter lattice.}
\begin{equation}
\Psi_0(\sigma, B) = \delta_{B,0}\frac{\prod_{w\in\cR}\frac1{\sqrt{2\pi}}\Gamma(\frac12 - iw\cdot\sigma)}{\prod_{\alpha\in\Phi}\frac1{\sqrt{2\pi}}\Gamma(1-i\alpha\cdot\sigma)}=\delta_{B,0}\frac{\prod_{w\in\cR}\frac1{\sqrt{2\pi}}\Gamma(\frac12 - iw\cdot\sigma)}{\prod_{\alpha\in\Phi_+}\frac1{2\pi}\frac{\pi\alpha\cdot\sigma}{\sinh \pi\alpha\cdot\sigma}}.
\end{equation}
The shift operators algebra $\cA_C[\cT]$ can act on this function, generating a certain $\cA_C[\cT]$-module. This module has a bilinear form defined by
\begin{align}
&(\Psi_1(\sigma, B), \Psi_2(\sigma, B)) = \sum_{B\in\Lambda^\vee}\int_{\mathfrak{t}}\dd^{{\rm rk}}\sigma\, \mu(\sigma,B) \Psi_1(\sigma,B)\Psi_2(\sigma,B),\cr 
&\mu(\sigma,B)=\prod_{\alpha\in\Phi_+}(-1)^{\alpha\cdot B}\left[(\alpha\cdot\sigma)^2 + \left(\frac{\alpha\cdot B}{2} \right)^2 \right]\prod_{w\in\cR}(-1)^{\frac12(|w\cdot B|-w\cdot B)} \frac{\Gamma(\frac12 + iw\cdot\sigma + \frac12 |w\cdot B|)}{\Gamma(\frac12 -iw\cdot\sigma + \frac12|w\cdot B|)},\cr
\end{align}
and we construct the trace of $\cO\in\cA_C[\cT]$ according to
\begin{equation}
\label{psOps}
\theta(\cO) = (\Psi_0, \cO \Psi_0).
\end{equation}
Notice that for \eqref{psOps}, the simpler expression for $\mu(\sigma,0)$ is enough, there is no need to know $\mu(\sigma,B)$ for general $B$. Another remark that will be important later is that introducing the FI terms $\zeta$ can be formulated as a modification of $\mu(\sigma,0)$:
\begin{equation}
\mu(\sigma,0)\mapsto \mu(\sigma,0) e^{2\pi i \zeta(\sigma)},
\end{equation}
where we think of $\zeta$ as a character of $\mathfrak{g}$. The trace, still defined by the above formula, receives an additional twist due to the FI terms.

Coupling to the bulk promotes masses to dynamical variables, which centrally extends the algebra and effectively undoes the central quotient mentioned above. Therefore, equations \eqref{Coul_Yang} describe yet another homomorphism 
\begin{equation}
\cY[\mathfrak{gl}_n] \to \cA^{(s)}_{N;n},
\end{equation}
where we now think of $\cA^{(s)}_{N;n}$ as the quantum Coulomb branch algebra of the NS5 interface.

Notice that the Coulomb branch perspective naturally leads to the Drinfeld's second presentation of the Yangian \cite{Drinfeld:1987sy}, as described above. On the other hand, the Higgs branch perspective of the previous section naturally lead to the RTT presentation. The relation between the two is quite non-trivial, involving quantum minors (see e.g. \cite[Theorem 12.1.4]{GuideToQu}), which we will use in the simplest example below. For now, we note that the embedding found using the Coulomb branch perspective does not have to match the embeddings in RTT presentations found in the previous section, but rather may differ by some automorphisms. A precise match across S-duality requires identification of the latter, but we do not do it here.

\paragraph{Coproduct.} We should mention another property of the algebras that has been made manifest in the Coulomb branch perspective: a coproduct $\cY_0^{\lambda_1+\lambda_2}[\mathfrak{sl}_n]\to \cY_0^{\lambda_1}[\mathfrak{sl}_n]\otimes \cY_0^{\lambda_2}[\mathfrak{sl}_n]$ compatible with the Yangian coproduct \cite{WeekesThesis} (it corresponds to splitting flavors and colors of the quiver into two subquivers while preserving the balanced condition). Intuitively, the defect associated to a collection of 't Hooft lines can be related to the fusion of two defects associated to sub-collections. 

For the $\cY[\mathfrak{gl}_{n}]$ Yangian, the coproduct takes the simple form: 
\begin{equation}
T[u;N+N']^a_c \to T[u;N]^a_b T[u;N']^b_c.
\end{equation}
We also expect the coproduct property to hold for the interface algebras as well. This is compatible with the expectation that $Q_0(u) T[u]$ and $Q_n(u) \bar T[u]$ should be polynomials in $u$,
as the $Q_0(u)$ and $Q_n(u)$ factor in the same way as the $T[u]$ matrices. A generic algebra can be obtained from the coproduct of 
simpler building blocks, such as ${\cal A}_{1;n}$ and ${\cal A}^{(1)}_{0;n}$, for which we demonstrated explicitly (here and in the companion paper) on the Higgs side
that $Q_0(u) T[u]$ and $Q_n(u) \bar T[u]$ are polynomails in $u$. 

One interesting consequence of the coproduct is that the space of possible traces on the algebras acquires a multiplication operation: 
we can define a trace on $\cY_0^{\lambda_1+\lambda_2}[\mathfrak{sl}_n]$ by mapping to $\cY_0^{\lambda_1}[\mathfrak{sl}_n]\otimes \cY_0^{\lambda_2}[\mathfrak{sl}_n]$
and taking independent traces of the two factors. We will discover in examples that  $\theta(\cO)$ admits a decomposition as a sum of such products of simpler traces. 

In fact, we will base our derivation of traces partially on the coproduct property. A general $\cA^{(s)}_{N;n}[m]$ admits coproduct maps into the tensor product of $N$ copies of $\cA_{1;n}[m]$ and $s$ copies of $\cA^{(1)}_{0;n}[m]$. Traces on $\cA_{1;n}[m]$ and $\cA^{(1)}_{0;n}[m]$ produce traces on $\cA^{(s)}_{N;n}[m]$ via coproduct maps. There are many such maps corresponding to the different ways of distributing masses entering $\cA^{(s)}_{N;n}[m]$ among the building blocks $\cA_{1,n}[m]$ and $\cA^{(1)}_{0;n}[m]$, each producing a possibly different trace. In the example $\cA_{N;2}[m]$ that we consider later, we will be able to generate all traces in this way.

\section{Traces and transfer matrices}\label{sec:traces}
We now turn to the question of twisted traces on $\cA^{(s)}_{N;n}$, which encode the data of sphere correlators on the interface. The setting is that of a 4d $\cN=4$ SYM on $S^4$, with the NS5 interface splitting it in two equal halves. The gauge group is $U(N)$ on one hemisphere and $U(N+ns)$ on the other.

Because of the morphism $\cY[\mathfrak{gl}_n]\to \cA^{(s)}_{N;n}$, any trace on $\cA^{(s)}_{N;n}$ will induce a trace on the Yangian $\cY[\mathfrak{gl}_n]$. Knowledge of the latter is enough to determine a large class of correlators, even all of them whenever the map $\cY[\mathfrak{gl}_n]\to \cA^{(s)}_{N;n}$ is surjective.

First, note the relation between traces on $\cY[\mathfrak{gl}_n]$ and $\cY[\mathfrak{sl}_n]$. Let us denote the twisted trace on the latter (induced by a map to $\cA^{(s)}_{N;n}[m]$) as $\theta_{m^L, m^R}$, where we explicitly indicate its dependence on masses, and keep dependence on FI parameters $\zeta$ (that determine the twist) implicit. Then the twisted trace $\theta$ on $\cY[\mathfrak{gl}_n]$ is given by coupling $\theta_{m^L, m^R}$ to the bulk:
\begin{equation}
\label{trace_bulk_interf}
\theta(\cO) = \frac1{|\cW|}\int_{\R^N \times \R^{N+ns}} [\dd m^L][\dd m^R] \underbrace{e^{-\frac{i}{\tau}\tr(m^L)^2}\Delta(m^L)\mathbbm{\Delta}(m^L)}_{\text{left hemisphere}} \theta_{m^L,m^R}(\cO) \underbrace{e^{-\frac{i}{\tau}\tr(m^R)^2}\Delta(m^R)\mathbbm{\Delta}(m^R)}_{\text{right hemisphere}},
\end{equation}
where $\Delta(a)$ and $\mathbbm{\Delta}(a)$ are the Vandermonde and the sinh-Vandermonde defined for a Lie algebra $\mathfrak{g}$, its positive root system $\Phi_+$, and $a\in\mathfrak{g}$ by
\begin{equation}
\Delta(a)=\prod_{\alpha\in\Phi_+}\langle\alpha,a\rangle,\quad \mathbbm{\Delta}(a)=\prod_{\alpha\in\Phi_+} 2\sinh\pi\langle\alpha,a \rangle.
\end{equation}
Due to the mass-dependent projection $\cY[\mathfrak{gl}_n]\to\cY[\mathfrak{sl}_n]$, every trace on $\cY[\mathfrak{sl}_n]$ is automatically a trace on $\cY[\mathfrak{gl}_n]$, -- that is why we are able to evaluate $\theta_{m^L, m^R}$ on $\cO\in \cY[\mathfrak{gl}_n]$. After that, we take a continuous linear conbination of such $\theta_{m^L, m_R}$ as $(m^L, m^R)$ varies over $\R^N \times \R^{N+ns}$, which is the Cartan of $U(N)\times U(N+ns)$.

It remains to determine $\theta_{m^L, m^R}$. We know one crucial property: it should be a twisted trace for $\cA^{(s)}_{N;n}[m]$, which is the central quotient of $\cA^{(s)}_{N;n}$. The space of such traces is known to be finite-dimensional \cite{kamnitzer2018quantum}. In favourable circumstances, the dimensionality is essentially that of the number of massive vacua of the 3d theory \cite{Gaiotto:2019mmf} and we expect that a basis can be provided by analytic continuation of traces over Verma modules associated to the vacua \cite{Bullimore:2016hdc}. \footnote{Since the Verma modules are infinite-dimensional, one must keep the twist parameters generic so the traces remain convergent. Only very special linear combinations of Verma traces (for example, those encoding sphere correlators) allow to switch off the twist parameters.}
Furthermore, the coproducts $\cY_0^{\lambda_1+\lambda_2}[\mathfrak{sl}_n]\to \cY_0^{\lambda_1}[\mathfrak{sl}_n]\otimes \cY_0^{\lambda_2}[\mathfrak{sl}_n]$ from \cite{WeekesThesis} (which are compatible with the Yangian coproduct) allow one to produce a large collection 
of traces for $\cA^{(s)}_{N;n}[m]$ from elementary traces for  $\cA^{(0)}_{1;n}[m]$ and $\cA^{(1)}_{0;n}[m]$. Experimentally, this collection seems large enough to provide all traces for $\cA^{(s)}_{N;n}[m]$. \footnote{The principle that general traces can be built from simpler traces via the coproduct 
can be implemented directly for the Yangian, but infinite linear combinations may be needed. Factoring through the truncated Yangians makes the problem finite-dimensional.} 

The algebras $\cA^{(0)}_{1;n}[m]$ and $\cA^{(1)}_{0;n}[m]$ are all truncations of $U(\mathfrak{sl}_n)$, so the associated traces 
are actually built from Verma modules for $U(\mathfrak{sl}_n)$, promoted to Yangian modules by the basic map
\begin{equation}
\cY[\mathfrak{sl}_n] \to U(\mathfrak{sl}_n),\quad T[u] \mapsto 1 + \frac{t^{[1]}}{u}.
\end{equation}
Twisted traces over such modules are well-known objects, they give transfer matrices of some spin chains with twisted-periodic boundary conditions. 

To review these notions following \cite{Bazhanov:2010ts,Bazhanov:2010jq} (see also \cite{Frassek:2018try,FinTsy,frassek2020lax} for constructions of Lax matrices), fix a highest weight representation of $\mathfrak{sl}_n$ with the highest weight $j$ and the representation space $V_j$. Then one constructs an L-operator\footnote{Notice that we use the ``Wick-rotated'' convention for the spectral parameter, following \cite[Section 2 and Appendix B]{Bazhanov:2010ts}, but \emph{not} their Section 3.}:
\begin{equation}
\mathbb{L}(u)=u + i\sum_{ab} E_{ab}\otimes J_{ab},
\end{equation}
which is an $n\times n$ matrix with ${\rm End}(V_j)$-valued entries. Here $E_{ab}$ are elementary $\mathfrak{sl}_n$ matrices (generators in the fundamental representation), and $J_{ab}$ are the corresponding $\mathfrak{sl}_n$ generators in the representation $V_j$. Because $J_{ab}$ obey the $\mathfrak{sl}_n$ commutation relations, one can show that $\mathbb{L}(u)$ obeys the RLL relation with the rational R-matrix. Namely, with
\begin{equation}
R(u) = u + iP: \C^n\otimes \C^n \to \C^n \otimes \C^n,
\end{equation}
where $P$ permutes the two $\C^n$'s, the following holds
\begin{equation}
R(u-v)(\mathbb{L}(u)\otimes \mathbbm{1})(\mathbbm{1}\otimes \mathbb{L}(v))=(\mathbbm{1}\otimes \mathbb{L}(v))(\mathbb{L}(u)\otimes \mathbbm{1})R(u-v).
\end{equation}
Assigning a spectral parameter $x$ to $V_j$, the evaluation module is determined by $\mathbb{L}(u)$ and defined on generators by
\begin{equation}
T[u] \mapsto \frac1{u}\mathbb{L}(u-x).
\end{equation}
It is common to denote $\mathbb{L}(u-x)$ pictorially as a crossing of two lines:
\begin{center}
	\includegraphics[scale=1]{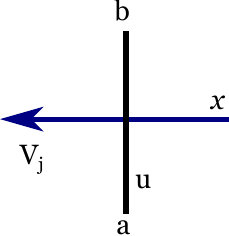}
\end{center}
where the fundamental representation runs along the vertical line, and that line also carries a spectral parameter $u$, while the representation $V_j$ runs along the horizontal line carrying the spectral parameter $x$. In these pictorial notations, the product $T[u_1]^{a_1}_{b_1} T[u_2]^{a_2}_{b_2}\dots T[u_L]^{a_L}_{b_L}$ acting in the evaluation module corresponding to the $\mathfrak{sl}_n$-module $V_j$ can be represented by
\begin{center}
	\includegraphics[scale=1]{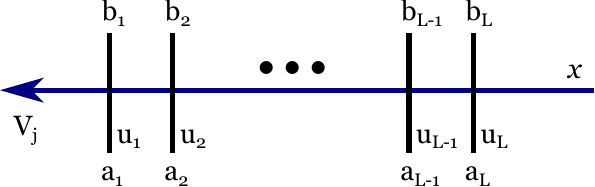}
\end{center}
while the twisted trace over this evaluation module is denoted by closing the horizontal line:
\begin{center}
	\includegraphics[scale=1]{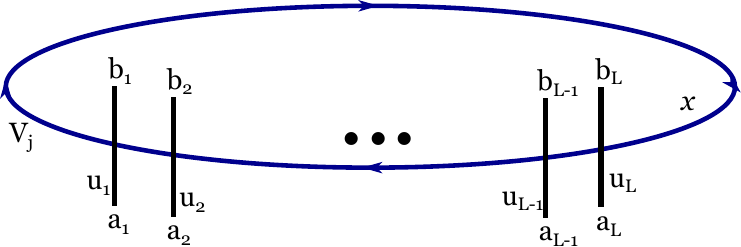}
\end{center}
We do not explicitly indicate the twist on such pictures, but it must be non-zero in order for traces over possibly infinite-dimensional modules to make sense. The above picture is quite familiar in integrability: it represents the transfer matrix of the $\mathfrak{sl}_n$ spin chain with $L$ sites, twisted-periodic boundary conditions, and impurities $u_i$. The transfer matrix corresponding to a finite-dimensional $\mathfrak{sl}_n$-module $V_j$ of spectral parameter $x$ is denoted by $\mathbb{T}_j(\vec{u}-x \vec{e})$, while if $V_j$ is a Verma module, it is called $\mathbb{T}_j^+(\vec{u}-x\vec{e})$, where $\vec{u}=(u_1,u_2,\dots,u_L)$ and $\vec{e}=(1,1,\dots,1)$.  We therefore find that the twisted trace of $T[u_1]^{a_1}_{b_1} T[u_2]^{a_2}_{b_2}\dots T[u_L]^{a_L}_{b_L}$ over the evaluation module corresponding to the Verma module $V_j$ is given by the matrix element of the transfer matrix:
\begin{equation}
\mathbb{T}^+_j(\vec{u}-x\vec{e})^{a_1\dots a_L}_{b_1\dots b_L},
\end{equation}
between the basis states of $\underbrace{\C^n\otimes \dots \otimes\C^n}_{L}$, which is the Hilbert space of the spin chain. If $V_j$ is the finite-dimensional $\mathfrak{sl}_n$ module, we likewise look at the matrix element of $\mathbb{T}_j(\vec{u}-x\vec{e})$.

As mentioned earlier, the twisted traces on $\cY[\mathfrak{sl}_n]$ we need to consider can be generated algebraically from twisted traces over the evaluation modules. For the $\mathfrak{sl}_n$-modules $V_{j_1}$, $V_{j_2}$, $\dots$, $V_{j_m}$, denote the evaluation modules by $\cV_{j_1},\dots, \cV_{j_m}$, and their spectral parameters by $x_1,\dots,x_m$. We may consider a linear span of traces on the tensor products of $\cV_j$'s,
\begin{equation}
\label{tr_over_tens_prod}
\tr_{\cV_{j_1}\otimes \dots \otimes \cV_{j_m}}\left( e^{-2\pi \zeta\cdot J} T[u_1]^{a_1}_{b_1} \dots T[u_L]^{a_L}_{b_L} \right).
\end{equation}
In this expression, the exponential factor $e^{-2\pi\zeta\cdot J}$ describes the twist, with $J\in\mathfrak{sl}_n$ and $\zeta$ from the Cartan subalgebra of $\mathfrak{sl}_n$. The tensor product of modules makes sense due to the existence of coproduct, implying that $T[u]^a_b$ acts on $\cV_{j_1}\otimes \dots \otimes \cV_{j_m}$ via
\begin{equation}
\sum_{c_1,\dots,c_{m-1}}T[u]^a_{c_1}\otimes T[u]^{c_1}_{c_2}\otimes \dots \otimes T[u]^{c_{m-1}}_b.
\end{equation}
This justifies the following pictorial representation of \eqref{tr_over_tens_prod}:
\begin{center}
	\includegraphics[scale=1]{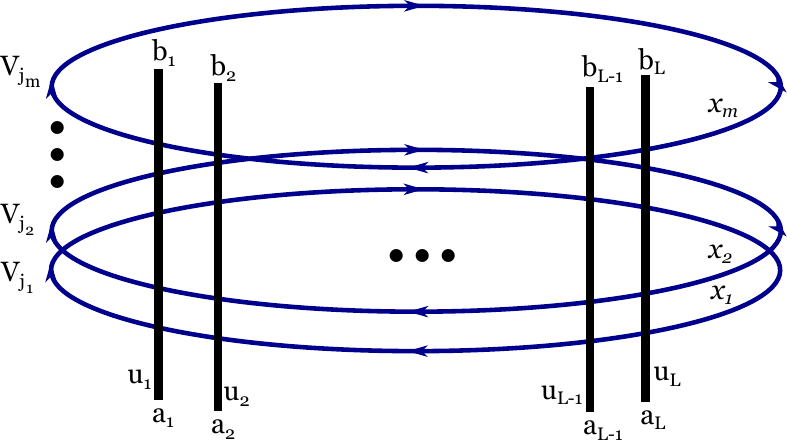}
\end{center}
Each horizontal loop here represents a transfer matrix, and stacking them on top of each other simply means taking a product of transfer matrices. Assuming for definiteness that all the $V_j$'s are Verma modules (of highest weights $j$), the corresponding transfer matrices are $\mathbb{T}^+_j(\vec{u})$, and the above picture encodes the following expression:
\begin{equation}
\label{prod_of_T_mat}
\langle a_1, \dots, a_L| \prod_{k=1}^m \mathbb{T}^+_{j_k}(\vec{u}-x_k\vec{e})|b_1,\dots, b_L\rangle,
\end{equation}
where we use the bra and ket notations for the matrix elements between states in $(\C^n)^{\otimes L}$, which is the Hilbert space of the length-$L$ $\mathfrak{sl}_n$ spin chain. Because all transfer matrices commute with each other, we can take their product in any order in the above expression. Furthermore, they can be simultaneously diagonalized by the Bethe eigenstates, in which case $\langle A| \prod_{k=1}^m \mathbb{T}^+_{j_k}(\vec{u}-x_k\vec{e})|B\rangle$ is the same as $\prod_{k=1}^m \langle A| \mathbb{T}^+_{j_k}(\vec{u}-x_k\vec{e})|B\rangle$, where $A$ and $B$ are some unit-norm eigenstates. We do not diagonalize transfer matrices, as for our purposes the standard basis on $(\C^n)^{\otimes L}$ is good enough.

We are thus left with the problem of determining the correct linear combination of the products of $\cA^{(0)}_{1;n}[m]$ and $\cA^{(1)}_{0;n}[m]$
transfer matrices which reproduce the 3d sphere protected correlation functions. This is a straightforward, if somewhat tedious, combinatorial problem. The Verma module traces come with very specific exponential prefactors involving bilinears of masses and FI parameters. 
These prefactors encode the weight of the highest weight vectors in the modules and essentially identify them uniquely when the masses are generic. 

As a consequence, the coefficients of the linear combinations of elementary traces can essentially be read off from the expansion of the 
partition function (with no operators inserted) into linear combinations of Verma module characters. In turn, that expansion can be derived from 
a standard sum-over-residues evaluation of the localization formula for the partition function, as in \cite{Gaiotto:2019mmf}.
Correlators with some simple insertions can be used as an extra check of the final formula. 

\subsection{The case of $\cA_{N;2}$ and $\cY[\mathfrak{gl}_2]$}\label{sec:XXX}
Let us solve this in one of the simplest cases, which is the algebra $\cA_{N;2}$ living at the intersection of $N$ D3 branes and $2$ fivebranes; we use the NS5 description in what follows. As we know, $\cA_{N;2}$ admits a surjective morphism from $\cY[\mathfrak{gl}_2]$, but first we have to look at the purely 3d theory at the interface, whose algebra admits a morphism from $\cY[\mathfrak{sl}_2]$. The constructions described above provide us with three generating series of shift operators: $F(z), E(z), H(z)$, which obey Drinfeld's second definition of the Yangian $\cY[\mathfrak{sl}_2]$. Using \cite[Theorem 12.1.4]{GuideToQu}, we can translate them into the RTT presentation according to
\begin{align}
\label{RTT_to_Dr}
F(z) &= T[z]^1_2(T[z]^1_1)^{-1},\quad E(z)=(T[z]^1_1)^{-1}T[z]^2_1,\cr
H(z)&= (T[z]^1_1)^{-1}(T[z+\epsilon]^1_1)^{-1} {\rm qdet}\, T[z+ \epsilon/2].
\end{align}
In \cite{Gerasimov_2005} a presentation in terms of certain generating functions $(A_i(z), B_i(z), C_i(z), D_i(z))$ was also worked out, and in the case of $\mathfrak{sl}_2$, it coincides with the above RTT presentation (for $n>2$, these generating functions are related to quantum minors of $T[u]$). We read off expressions for the generators $T[u]^a_b$ from \eqref{RTT_to_Dr}:
\begin{align}
\label{RTT_sl2}
T[u]^1_1 &= \prod_{a=1}^N \left(1 - \frac{\phi_{1,a}}{u} \right),\cr
T[u]^1_2 &= \sum_{a=1}^N \frac{u^+_{1,a}}{u}\prod_{b\neq a}\left(1-\frac{\phi_{1,b}}{u} \right),\cr
T[u]^2_1 &= \sum_{a=1}^N\frac{u^-_{1,a}}{u}\prod_{b\neq a}\left(1-\frac{\phi_{1,b}}{u} \right),\cr
T[u]^2_2&=T[u]^1_1H(u)+T[u]^1_2(T[u]^1_1)^{-1}T[u]^2_1\cr
&=\frac{Q_0(u+\frac{\epsilon}{2})Q_2(u+\frac{\epsilon}{2})}{u^N Q_1(u+\epsilon)} + \left(\prod_{a=1}^N (1-\frac{\phi_a}{u}) \right)\left(\sum_{a=1}^N \frac1{u-\phi_a}u^+_a \right)\left(\sum_{a=1}^N \frac1{u-\phi_a}u^-_a \right).\cr
\end{align}
We can also check that (compare to \cite[Section 2.2]{frassek2020lax}):
\begin{equation}
{\rm qdet}\, T[u+\epsilon/2] = \prod_{a=1}^N \left(1 + \frac{\frac{\epsilon}{2} - m^L_a}{u} \right)\left(1-\frac{\frac{\epsilon}{2}+m^R_a}{u+\epsilon} \right),
\end{equation}
which is indeed a formal series in $u$ with numerical coefficients determined by masses, as was stated before.

The above expressions for $T[u]^a_b$ are not unique, we can apply an automorphism
\begin{equation}
T[u] \mapsto f(u)T[u],\, \text{with}\, f(u)\in 1 + u^{-1}\C[u^{-1}],
\end{equation}
which will modify $T[u]$ without affecting the expressions for $F(z)$, $E(z)$, and $H(z)$. Other two known automorphisms of the RTT presentation are:
\begin{align}
T[u] &\mapsto T[u+c],\, c\in\C,\cr
T[u] &\mapsto B T[u] B^{-1},\, B\in SL(n,\C) \,\, (\text{which in our case is } SL(2,\C)),
\end{align}
and we expect that identification of the generators in \eqref{RTT_sl2} with those in the S-dual D5 construction of Section \ref{sec:N_cross_n} involves a certain combination of all three automorphisms.

Despite ambiguity in choosing the RTT presentation, \eqref{RTT_sl2} provides a valid set of generators, and we are simply going to work with it. In particular, we will compute some correlators in order to determine which modules appear in the product of transfer matrices in the equation \eqref{prod_of_T_mat}.

The 3d theory on the interface in the $\cY[\mathfrak{sl}_2]$ case is simply the $U(N)$ SQCD with $2N$ fundamental flavors, but coupling to the bulk suggests to break them into a group of $N$ fundamentals and a group of $N$ anti-fundamentals, and rather think of the quiver:
\begin{center}
	\includegraphics[scale=0.6]{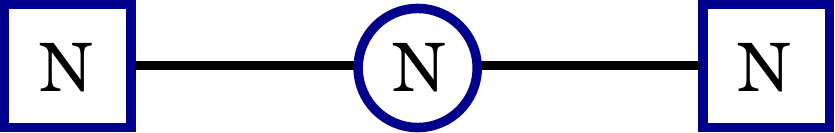}
\end{center}
We analyze this theory in the Appendix \ref{app:SQCD}, describing its partition function and correlators. In particular, the partition function is, up to a sign, as in \cite{Gaiotto:2019mmf},
\begin{equation}
\label{Z_SQCD_sum}
Z=\sum_{\sigma\in\substack{\frac{S_{2N}}{S_{N}\times S_N}}} \frac{i^{-N^2}\prod_{a=1}^N e^{2\pi i\zeta\mu_{\sigma(a)}}}{(e^{\pi\zeta}-e^{-\pi\zeta})^N \prod_{a=1}^N \prod_{k=N+1}^{2N} 2\sinh\pi(\mu_{\sigma(a)} - \mu_{\sigma(k)})},
\end{equation}
where the dimensionless left and right masses are again grouped into a single vector,
\begin{equation}
(\mu_1, \mu_2,\dots, \mu_{2N})= (\ell m^L_1,\dots,\ell m^L_N, \ell m^R_1,\dots, \ell m^R_N),
\end{equation}
and the sum runs over the choices of $N$-tuples of masses (i.e., over the ${2N\choose N}$ massive vacua). When coupling to the bulk, it will be important to distinguish left and right masses again.

It is argued in the Appendix \ref{app:SQCD} that the correlation functions of Yangian generators are captured in terms of the matrix elements of transfer matrices, as anticipated above. Namely, the answer takes the form:
\begin{align}
\label{answer_3d_prel}
u_1^{N}\dots u_L^N&\langle T[u_1]^{a_1}_{b_1}\dots T[u_L]^{a_L}_{b_L} \rangle \cr
=\sum_{\sigma\in\substack{\frac{S_{2N}}{S_{N}\times S_N}}} &\frac{i^{-N^2} e^{i\pi\zeta\sum_{j=1}^{2N}\mu_j} }{\prod_{a=1}^N \prod_{k=N+1}^{2N}2\sinh \pi(\mu_{\sigma(a)}-\mu_{\sigma(k)})}\cr 
&\times\Big\langle a_1,\dots,a_L;L\Big|\prod_{a=1}^N \mathbb{T}^+_{-\frac12 +i\frac{\mu_{\sigma(a)}-\mu_{\sigma(a+N)}}{2}}\left(\vec{z}-\frac{\mu_{\sigma(a)}+ \mu_{\sigma(a+N)}}{2}\vec{e}\right)\Big|b_1,\dots,b_L;L\Big\rangle,\cr
\end{align}
where we use the notation $\vec{e}=(\underbrace{1,\dots,1}_L)$. Here $\mathbb{T}^+_j(\vec{u})$ is the transfer matrix of a length-$L$ inhomogeneous XXX spin chain, for the evaluation module based on the $\mathfrak{sl}_2$ Verma module of highest weight $j$. We compute the matrix elements of a product of transfer matrices between the basis vectors of the spin chain Hilbert space $(\C^2)^{\otimes L}$. In the vectors $|a_1,\dots,a_L;L\rangle$, all the $a_i$'s take values $1$ and $2$. We may think of $|1,\dots,1,1;L\rangle$ as $|\uparrow\dots \uparrow\uparrow\rangle$, of $|1,\dots,1,2;L\rangle$ as $|\uparrow\dots \uparrow\downarrow\rangle$, etc.

The values $j_a= -\frac12 +i\frac{\mu_{\sigma(a)}-\mu_{\sigma(a+N)}}{2}$ of highest weights, and the spectral parameters $x_a = \frac{\mu_{\sigma(a)}+\mu_{\sigma(a+N)}}{2}$ of the evaluation modules entering \eqref{answer_3d_prel} deserve an explanation. For that, recall the relation between the transfer matrix $\mathbb{T}^+_j(\vec{u})$ and the Baxter Q-operators \cite{Bazhanov:1996dr,Bazhanov:2010ts}:
\begin{equation}
\mathbb{T}^+_j(\vec{u})=\frac1{2i\sin\frac{\phi}{2}} \mathbf{Q}_+\left(\vec{u} +i(j+1/2) \vec{e}  \right)\mathbf{Q}_-\left(\vec{u} -i(j+1/2)\vec{e} \right),
\end{equation}
where $\phi=-2\pi i \zeta$ is our twist parameter, and $\mathbf{Q}_\pm$ can be computed as traces of monodromy matrices over an auxiliary oscillator Fock space following \cite{Bazhanov:2010ts}. In terms of these we find:
\begin{align}
\label{answer_3d}
u_1^{N}\dots u_L^N&\langle T[u_1]^{a_1}_{b_1}\dots T[u_L]^{a_L}_{b_L} \rangle \cr
=\sum_{\sigma\in\substack{\frac{S_{2N}}{S_{N}\times S_N}}} &\frac{i^{-N^2} e^{i\pi\zeta\sum_{j=1}^{2N}\mu_j} }{(2\sinh \pi\zeta)^N\prod_{a=1}^N \prod_{k=N+1}^{2N}2\sinh \pi(\mu_{\sigma(a)}-\mu_{\sigma(k)})}\cr 
&\times\Big\langle a_1,\dots,a_L;L\Big|\prod_{a=1}^N \mathbf{Q}_+(\vec{u}-\mu_{\sigma(a)}\vec{e})\mathbf{Q}_-(\vec{u}-\mu_{\sigma(a+N)}\vec{e})\Big|b_1,\dots,b_L;L\Big\rangle.\cr
\end{align}
Recall that the Q-operators commute with each other and with the transfer matrices. Thus this expression, unlike \eqref{answer_3d_prel}, makes symmetry under the permutations $\sigma\in S_N\times S_N$ manifest, showing that there are indeed only ${2N\choose N}$ nontrivial terms in the above sum.

More importantly, \eqref{answer_3d} makes connection with how we derive it in the Appendix \ref{app:SQCD}. The algebra $\cA_{N;2}[m]$ (the central quotient of $\cA_{N;2}$) admits a coproduct map into $\cA_{1;2}[m]^{\otimes N}$, which corresponds to breaking up $U(N)$ SQCD with $2N$ flavors into $N$ instances of $U(1)$ SQED with 2 flavors. In the process, each SQED inherits two masses from the SQCD, so we have to split $2N$ masses $\mu_1,\dots,\mu_{2N}$ into $N$ pairs. If we look at the copy of SQED with masses $\mu_a$ and $\mu_b$, one can explicitly find that it has two traces corresponding to Verma modules (see Appendix \ref{app:SQCD}), given by
\begin{equation}
\label{sqed_trace}
\mathbf{Q}_+(\vec{u} - \mu_a \vec{e})\mathbf{Q}_-(\vec{u}-\mu_b \vec{e})\quad \text{and} \quad \mathbf{Q}_+(\vec{u} - \mu_b \vec{e})\mathbf{Q}_-(\vec{u}-\mu_a \vec{e})
\end{equation}
in the above formula \eqref{answer_3d}. Using the coproduct map $\cA_{N;2}[m]\to \cA_{1;2}[m]\otimes\dots\otimes\cA_{1;2}[m]$ and the choice of trace on each $\cA_{1;2}[m]$, we can build a trace on $\cA_{N;2}[m]$. The resulting trace is a product of $N$ traces like \eqref{sqed_trace}, so we conclude that the trace on $\cA_{N;2}[m]$ corresponds to a product of Q-operators 
\begin{equation}
\prod_{a=1}^N \mathbf{Q}_+(\vec{u}-\mu_{\sigma(a)}\vec{e})\mathbf{Q}_-(\vec{u}-\mu_{\sigma(a+N)}\vec{e}),
\end{equation}
for some permutation $\sigma$ of $2N$ masses. There are precisely ${2N\choose N}$ such traces, the same as the number of vacua. These are simply traces over the Verma modules of the Coulomb branch algeba of SQCD (which can be clearly seen at generic masses, where such modules are irreducible). There are ${2N\choose N}$ such Verma modules, in correspondence with massive vacua \cite{Bullimore:2016hdc,Gaiotto:2019mmf}, and we thus have found them all. They generate all traces \cite{Etingof:2019guc}, and to find the trace computing sphere correlators, we have to identify their correct linear combination. We find in the Appendix \ref{app:SQCD} that it is given by \eqref{answer_3d}.

Now we couple this to the bulk by inserting \eqref{answer_3d} into \eqref{trace_bulk_interf} and integrating over masses. The masses $(\mu_1,\dots,\mu_N)\equiv (\mu_1^L,\dots,\mu_N^L)$ are coupled to the left and called ``left masses'', similarly $(\mu_{N+1},\dots,\mu_{2N})=(\mu_1^R,\dots,\mu_N^R)$ are the right masses (though the precise choice of left-right splitting is immaterial). In doing this, of great help is the following identity:
\begin{align}
\label{comb_id}
\sum_{\sigma\in\frac{S_{2N}}{S_N\times S_N}} \frac{\mathbbm{\Delta}(\mu^L)\mathbbm{\Delta}(\mu^R)}{\prod_{a=1}^N \prod_{k=N+1}^{2N}2\sinh \pi(\mu_{\sigma(a)}-\mu_{\sigma(k)})} \prod_{a=1}^N \mathbf{Q}_+(\vec{u}-\mu_{\sigma(a)}\vec{e})\mathbf{Q}_-(\vec{u}-\mu_{\sigma(a+N)}\vec{e})\cr
=(-1)^{\frac{N(N-1)}{2}}\sum_{s\in S_N}(-1)^s \prod_{a=1}^N \frac{\mathbf{Q}_+(\vec{u}-\mu^L_a\vec{e})\mathbf{Q}_-(\vec{u}-\mu^R_{s(a)}\vec{e})-\mathbf{Q}_-(\vec{u}-\mu^L_a\vec{e})\mathbf{Q}_+(\vec{u}-\mu^R_{s(a)}\vec{e})}{2\sinh\pi(\mu^L_a-\mu^R_{s(a)})},
\end{align}
which we prove in the Appendix \ref{app:comb} using commutativity of $\mathbf{Q}_\pm$ and some combinatorics.

We apply the identity \eqref{comb_id} to rearrange the Q-operators in \eqref{answer_3d}, insert the result in the integral \eqref{trace_bulk_interf}, and recognize that the bulk contribution is permutation-invariant, to arrive at a much simpler matrix model. We also convert combinations of Q-operators under the product in \eqref{comb_id} into transfer matrices of Verma modules, resulting in
\begin{align}
\label{answer_4d}
u_1^{N}\dots u_L^N&\langle T[u_1]^{a_1}_{b_1}\dots T[u_L]^{a_L}_{b_L} \rangle\cr
=\frac1{N!}\int_{\R^N\times\R^N}&[\dd \mu^L][\dd \mu^R] e^{-\frac{i\pi}{\tau}\tr(\mu^L)^2 - \frac{i\pi}{\tau}\tr(\mu^R)^2+2\pi i\zeta\sum_{a=1}^N \bar\mu_a}\Delta(\mu^L)\Delta(\mu^R)\cr &\times\Big\langle a_1,\dots,a_L;L\Big|\prod_{j=1}^N \frac{\mathbb{T}^+_{-\frac12 +i(\mu^L_j-\bar{\mu}_j)}(\vec{u}-\bar{\mu}_j\vec{e})-\mathbb{T}^+_{-\frac12 +i(\mu^R_j-\bar{\mu}_j)}(\vec{u}-\bar{\mu}_j\vec{e})}{2i\sinh\pi(\mu^L_j-\mu^R_j)}\Big|b_1,\dots,b_L;L\Big\rangle,\cr
\end{align}
where we have introduced the left-right averaged masses for brevity:
\begin{equation}
\bar\mu_j = \frac12(\mu^L_j + \mu^R_j).
\end{equation}
The expression \eqref{answer_4d} also encodes the interface partition function, which formally corresponds to the length $L=0$ spin chain with the transfer matrix:
\begin{equation}
\mathbb{T}^+_j= \frac{e^{-2\pi i \zeta(j+\frac12)}}{2 \sinh \pi\zeta}.
\end{equation}
In the Appendix \ref{app:SQCD} we also provide an alternative derivation, where instead of using \eqref{comb_id}, we go back to the integral expression for the 3d partition function, and after some further manipulations, we end up with the same answer.

Notice also that the combination of transfer matrices that we see in \eqref{answer_4d} is of the form $\mathbb{T}^+_{j}(\vec{u}) - \mathbb{T}^+_{-1-j}(\vec{u})$. For $j\in\frac12\Z$, this is precisely the transfer matrix $\mathbb{T}_j(\vec{u})$ of the irreducible finite-dimensional spin-$j$ module of $\mathfrak{sl}_2$. It admits the twist-zero limit $\zeta\to 0$. For general $j$, it does not have this interpretation, but still stays finite in the twist-zero limit \cite{Bazhanov:2010ts}, making this expected property of \eqref{answer_4d} manifest.

We see that while the answer in \eqref{answer_3d} for the 3d theory is remarkably explicit and algorithmic (when combined with the results of \cite{Bazhanov:2010ts} for the Q operators), the answer \eqref{answer_4d} for the interface is more involved, as one still has to compute a double-matrix integral. It is conceivable, however, that the expression \eqref{answer_4d} is amenable to the large-N analysis.

\appendix
\section{Yangian generators in $\cA_{N;n}$}\label{app:yang}
We have weight-$2$ generators $B_+$ and $B_-$ of $U(\mathfrak{gl}_N) \times U(\mathfrak{gl}_N)$, weight-$1$ generators $X^a$ and $Y_a$ of $\mathrm{Weyl}^{Nn}$, and a weight-2 quantization parameter $\hbar$. When possible, we will suppress $\mathfrak{gl}_N$ indices. If they are needed, we will use lowercase Greek indices. 

We have commutators
\begin{align}
[X^a_\alpha,Y_b^\beta] &= \hbar \delta^a_b \delta_\alpha^\beta  \cr
[(Y_a X^a)^\alpha_\beta, (Y_b X^b)^\gamma_\delta ] &= \hbar \delta^\gamma_\beta (Y_a X^a)^\alpha_\delta - \hbar \delta^\alpha_\delta (Y_a X^a)^\gamma_\beta \cr
[(B_+)^\alpha_\beta, (B_+)^\gamma_\delta ] &=\hbar \delta^\gamma_\beta (B_+)^\alpha_\delta - \hbar \delta^\alpha_\delta (B_+)^\gamma_\beta \cr
[(B_-)^\alpha_\beta, (B_-)^\gamma_\delta ] &= \hbar \delta^\gamma_\beta (B_-)^\alpha_\delta - \hbar \delta^\alpha_\delta (B_-)^\gamma_\beta
\end{align}

The quantum Hamiltonian reduction consists of taking quotient by the left ideal generated by the F-term relation 
\begin{equation}
\mu^\alpha_\beta \equiv (B_+)^\alpha_\beta + (B_-)^\alpha_\beta + Y_a^\alpha X^a_\beta=- \hbar N \delta^\alpha_\beta,
\end{equation}
and then restricting to the $\mathfrak{gl}_N$ invariant operators. The FI parameter can be absorbed into the diagonal component of $B_\pm$. We fixed it to a convenient value.

The quantum Hamiltonian reduction preserves the global $SL(n)$ symmetry acting on the Weyl algebra, with infinitesimal generators 
\begin{equation}
X^a Y_b- \frac{1}{n} \delta^a_b X^c Y_c. 
\end{equation}
We also define $\mathfrak{gl}_n$ generators
\begin{equation}
t^{[1]}{}^a_b = -X^a Y_b. 
\end{equation}

Notice that the F-term relation should be applied to the very right (or to the very left, had we chosen to work with the right ideal, which leads to the isomorphic answer) of gauge-invariant operators. 
We will often need to deal with the simplification of gauge-invariant expressions of the form (here $O^\beta$ and $O'_\alpha$ are $\mathfrak{gl}_N$-fundamental and anti-fundamental operators respectively)
\begin{equation}
O'_\alpha \mu^\alpha_\beta O^\beta = \hbar N O'_\alpha O^\alpha + O'_\alpha O^\beta \mu^\alpha_\beta =0,
\end{equation}
giving, for example, 
\begin{equation}
O'_\alpha (B_-)^\alpha_\beta O^\beta + O'_\alpha (B_+)^\alpha_\beta O^\beta+ O'_\alpha  Y_a^\alpha X^a_\beta O^\beta =0.
\end{equation}

We find it useful to define two generating functions of open trace operators:
\begin{equation}
T[u]^a_b = \delta^a_b - X^a \frac{1}{u-B_+} Y_b  \qquad \qquad \bar{T}[u]^a_b = \delta^a_b + X^a \frac{1}{u+B_-} Y_b.
\end{equation}
We can take the product
\begin{equation}
T[u]^a_b \bar T[u]^b_c= \delta^a_c - X^a \frac{1}{u-B_+} Y_c + X^a \frac{1}{u+B_-} Y_c- X^a \frac{1}{u-B_+} Y_b X^b \frac{1}{u+B_-} Y_c,
\end{equation}
and use the F-term relation above:
\begin{equation}
T[u]^a_b \bar T[u]^b_c= \delta^a_c - X^a \frac{1}{u-B_+} Y_c + X^a \frac{1}{u+B_-} Y_c+ X^a \frac{1}{u-B_+} (B_+ + B_-) \frac{1}{u+B_-} Y_c,
\end{equation}
which simplifies dramatically to 
\begin{equation}
T[u]^a_b \bar T[u]^b_c= \delta^a_c.
\end{equation}

We can also readily compute a commutator: 
\begin{equation}
[T[u]^a_b, \bar T[w]^c_d] =- \hbar \delta^a_d X^c \frac{1}{w+B_-} \frac{1}{u-B_+} Y_b + \hbar \delta^c_b  X^a  \frac{1}{u-B_+} \frac{1}{w+B_-} Y_d,
\end{equation}
and use it to get 
\begin{equation}
[T[u]^a_b, T[w]^c_d] =T[w]^c_e \left[ \hbar \delta^a_f X^e \frac{1}{w+B_-} \frac{1}{u-B_+} Y_b - \hbar \delta^e_b  X^a  \frac{1}{u-B_+} \frac{1}{w+B_-} Y_f \right] T[w]^f_d,
\end{equation}
i.e. 
\begin{equation}
[T[u]^a_b, T[w]^c_d] =  \hbar T[w]^c_e X^e \frac{1}{w+B_-} \frac{1}{u-B_+} Y_b \bar T[w]^a_d - \hbar \bar T[w]^c_b  X^a  \frac{1}{u-B_+} \frac{1}{w+B_-} Y_f T[w]^f_d,
\end{equation}
which simplifies by the F-term relations to 
\begin{equation}
[T[u]^a_b, T[w]^c_d] =  \hbar X^c \frac{1}{w-B_+}\frac{1}{u-B_+} Y_b T[w]^a_d - \hbar T[w]^c_b  X^a  \frac{1}{u-B_+} \frac{1}{w-B_+} Y_d, 
\end{equation}
i.e. to 
\begin{equation}
[T[u]^a_b, T[w]^c_d] = \frac{1}{u-w} \hbar X^c \left[\frac{1}{w-B_+}-\frac{1}{u-B_+} \right] Y_b T[w]^a_d -\frac{1}{u-w}  \hbar T[w]^c_b  X^a  \left[\frac{1}{w-B_+}-\frac{1}{u-B_+} \right] Y_d, 
\end{equation}
and then to the $\cY[\mathfrak{gl}_n]$ Yangian relation
\begin{equation}
[ T[u]^a_b, T[w]^c_d] =  \hbar \frac{ T[w]^c_b  T[u]^a_d- T[u]^c_b T[w]^a_d}{u-w}.
\end{equation}

Next, consider a small variant of the calculation above:
\begin{equation}
T[u]^a_b \bar T[w]^b_c= \delta^a_c + (u-w) X^a \frac{1}{u-B_+} \frac{1}{w+B_-} Y_c,
\end{equation}
and take a trace over the $\mathfrak{gl}_n$ indices:
\begin{equation}
T[u]^a_b \bar T[w]^b_a=  n + (u-w) \Tr  \frac{1}{w+B_-} Y_a X^a \frac{1}{u-B_+} +n \hbar (u-w)\Tr  \frac{1}{u-B_+} \frac{1}{w+B_-}. 
\end{equation}

We should learn how to apply the moment map within a trace:
\begin{equation}
(O')^\gamma_\alpha \mu^\alpha_\beta O_\gamma^\beta = \hbar N O'_\alpha O^\alpha- \hbar (O')^\alpha_\alpha  O_\beta^\beta + O'_\alpha O^\beta \mu^\alpha_\beta =
\hbar (O')^\alpha_\alpha  O_\beta^\beta,
\end{equation}
so that 
\begin{equation}
\Tr O' B_+ O + \Tr O' B_- O=-\hbar \Tr O' \Tr O - \Tr O' Y_a X^a O, 
\end{equation}
to get:
\begin{equation}
T[u]^a_b \bar T[w]^b_a=  n -\hbar (u-w) \Tr  \frac{1}{w+B_-} \Tr \frac{1}{u-B_+} -(u-w) \Tr  \frac{1}{w+B_-} (B_+ + B_- -n \hbar) \frac{1}{u-B_+}.
\end{equation}
Specializing to $w = u - n \hbar$, we find a nice answer:
\begin{equation}
T[u]^a_b \bar T[u - n \hbar]^b_a=  n\left(1-\hbar \Tr \frac{1}{u-B_+}\right)\left(1+\hbar \Tr  \frac{1}{u - n \hbar+B_-}\right).
\end{equation}
An equivalent calculation gives 
\begin{equation}
\bar T[u]^a_b T[u + n \hbar]^b_a=  n\left(1+\hbar \Tr \frac{1}{u+B_-}\right)\left(1-\hbar \Tr  \frac{1}{u + n \hbar-B_+}\right).
\end{equation}

The expressions on the right hand side play an important role in the theory of $U(\mathfrak{gl}_N)$.
The Harish-Chandra isomorphism can be written explicitly as  
\begin{equation}
1-\hbar \Tr  \frac{1}{u -B_+} = \frac{P_+(u-\hbar)}{P_+(u)},
\end{equation}
where $P_+(u)$ is a degree $N$ monic polynomial whose coefficients are a natural basis for the center of $U(\mathfrak{gl}_N)$. 
We can similarly write 
\begin{equation}
1+\hbar \Tr \frac{1}{u+B_-} = \frac{P_-(u+\hbar)}{P_-(u)}.
\end{equation}
Then we have: 
\begin{equation}
T[u]^a_b \bar T[u - n \hbar]^b_a=  n\frac{P_+(u-\hbar)}{P_+(u)}\frac{P_-(u-(n-1)\hbar)}{P_-(u- n \hbar)},
\end{equation}
and
\begin{equation}
\bar T[u]^a_b T[u + n \hbar]^b_a=  n\frac{P_+(u+ (n-1) \hbar)}{P_+(u+ n \hbar)}\frac{P_-(u+\hbar)}{P_-(u)}. 
\end{equation}
The left hand side can be thought of as the ratio of quantum determinants of $T[u]^a_b$ evaluated at $u+ \frac{n-1}{2}\hbar$ and at $u+ \frac{n+1}{2}\hbar$. The formula is compatible with the identification of the quantum determinant of $T[u]^a_b$
being $\frac{P_-(u-\frac{n-1}{2}\hbar)}{P_+(u+ \frac{n-1}{2} \hbar)}$ and the quantum determinant of $\bar T[u]^a_b$ 
being $\frac{P_+(u+\frac{n-1}{2}\hbar)}{P_-(u- \frac{n-1}{2} \hbar)}$.

\section{Trace and transfer matrices for SQCD}\label{app:SQCD}
The partition function of the $U(N)$ SQCD with $2N$ flavors reads:
\begin{equation}
\label{CouInt}
Z=\frac1{N!}\int\dd^N\sigma\, e^{2\pi i \zeta \sum_{a=1}^N \sigma_a} \frac{\prod_{a<b}4\sinh^2\pi(\sigma_a-\sigma_b)}{\prod_{a=1}^N\left[\prod_{i=1}^N 2\cosh\pi(\sigma_a-\mu_i^L) \prod_{i=1}^N 2\cosh\pi(\sigma_a-\mu_i^R) \right] },
\end{equation}
where we use dimensionless masses $\mu^L_i=\ell m_i^L$ and $\mu^R_i=\ell m_i^R$. One computes this integral by assuming $\zeta>0$ and closing the contour in the upper half plane for each $\sigma_a$ \cite{Gaiotto:2019mmf}. The poles are located at
\begin{align}
\label{poles}
\sigma_a &= \mu^L_{j_a} + i\left(n_a + \frac12\right),\quad \text{for } a=1\dots p,\cr
\sigma_a&= \mu^R_{j_a} + i\left(n_a + \frac12\right),\quad \text{for } a=p+1\dots N,
\end{align}
where $\{j_1,\dots, j_p\}\subset \{1,\dots,N\}$ is a subset of $p$ flavors on the left node, and $\{j_{p+1},\dots,j_N\}\subset \{1,\dots,N\}$ is a subset of $N-p$ flavors on the right node. We should also sum over $0\leq p\leq N$. To make the following formulas more compact, it is useful to introduce a notation uniting all masses into a single $2N$-component vector $\mu_j$:
\begin{equation}
(\mu_1, \mu_2,\dots, \mu_{2N})= (\mu^L_1,\dots,\mu^L_N, \mu^R_1,\dots, \mu^R_N),
\end{equation}
then the choice of $p$ flavors on the left and $N-p$ flavors on the right (for all $p$) is given by a choice of $\{i_1,\dots, i_N\}\subset \{1,\dots, 2N\}$. Furthermore, we can extend this to a permutation $\sigma$ of $\{1,\dots, 2N\}$, such that $\sigma(k)=i_k$ for $1\leq k\leq N$. Then computing resides at the poles and summing over $n_a=0,1,\dots$, we find as in \cite{Gaiotto:2019mmf} (up to a sign)
\begin{equation}
\label{SQCD_Z}
Z=\sum_{\sigma\in\substack{\frac{S_{2N}}{S_{N}\times S_N}}} \frac{i^{-N^2}\prod_{a=1}^N e^{2\pi i\zeta\mu_{\sigma(a)}}}{(e^{\pi\zeta}-e^{-\pi\zeta})^N \prod_{a=1}^N \prod_{k=N+1}^{2N} 2\sinh\pi(\mu_{\sigma(a)} - \mu_{\sigma(k)})}.
\end{equation}
Now we would like to compute correlators. For that, let us first go back in the computation of \eqref{SQCD_Z} and spell out the step where we sum over $n_a$. When we compute the residues, the only $n_a$-dependent contribution comes from evaluating $e^{2\pi i\zeta\sum_a \sigma_a}$ at the poles \eqref{poles}. In general, there is also a sign $(-1)^{n_a N_f}$ coming from the other terms, but in our case it cancels because $N_f=2N$. Including the sum over $n_a$, we find:
\begin{align}
\label{sum_na}
\prod_{a=1}^N \sum_{n_a\geq 0} e^{2\pi i\zeta\left(\mu_{\sigma(a)} + i \left( n_a + \frac12 \right) \right) } =
\frac{\prod_{a=1}^N e^{2\pi i \zeta \mu_{\sigma(a)}}}{(e^{\pi\zeta}-e^{-\pi\zeta})^N},
\end{align}
which is part of the answer above.

Let us include insertions. We start with one-point functions. It is clear that
\begin{equation}
\langle T[u]^1_2 \rangle = \langle T[u]^2_1 \rangle=0,
\end{equation}
because the corresponding monopole operators have non-zero topological charge. The other two components of $T[u]^a_b$ yield non-trivial one-point functions. Let us compute:
\begin{equation}
\langle T[u]^1_1 \rangle = u^{-N} \left\langle \prod_{a=1}^N (u-\phi_a)\right\rangle,
\end{equation}
so we have to include $\prod_a (u-\sigma_a)$ under the integral \eqref{CouInt}, where for simplicity we have absorbed a factor of $\ell=\frac{i}{\epsilon}$ into $u$ to make it dimensionless, so that we can write $\sigma_a$ instead of $\sigma_a/\ell$. Such insertion under the integral does not change the poles, and only modifies the residues in an obvious way. The whole effect is to modify \eqref{sum_na} as follows:
\begin{equation}
\label{na_sum_T11}
\prod_{a=1}^N \left[\sum_{n_a\geq 0} e^{2\pi i\zeta\left(\mu_{\sigma(a)} + i \left( n_a + \frac12 \right) \right) } \left(u-\mu_{\sigma(a)} - i \left( n_a + \frac12 \right)\right)\right]=\frac{\prod_{a=1}^N e^{2\pi i \zeta \mu_{\sigma(a)}}}{(e^{\pi\zeta}-e^{-\pi\zeta})^N}\prod_{a}(u-\mu_{\sigma(a)}-\frac{i}2\coth\pi\zeta).
\end{equation}

Let us compare this to transfer matrices in the length $L=1$ spin chain. The transfer matrix $\mathbb{T}^+_j(u)$ for the Verma module $V_j$ is related to the Baxter Q-operators:
\begin{equation}
\mathbb{T}^+_j(u) = \frac1{2i\sin\frac{\phi}2} \mathbf{Q}_+(u+i(j+\frac12))\mathbf{Q}_-(u-i(j+\frac12)),
\end{equation}
where we use the ``Wick-rotated'' convention for spectral parameter, following Section 2 and Appendix B (rather than Section 3) of \cite{Bazhanov:2010ts}. The expressions for the $L=1$ operators $\mathbf{Q}_\pm$ can be found in the Appendix B of \cite{Bazhanov:2010ts}. They are $2\times 2$ matrices, with the 1-1 components given by:
\begin{equation}
\mathbf{Q}_-(u)_1^1 = e^{-\frac{\phi}2 u},\quad \mathbf{Q}_+(u)_1^1 = e^{\frac{\phi}{2} u}(u-\frac12 \cot\frac{\phi}{2}),
\end{equation}
so the 1-1 component of the transfer matrix (which is an already diagonal $2\times 2$ matrix) is
\begin{equation}
\mathbb{T}^+_j(u)_1^1=\frac{e^{i\phi(j+\frac12)}}{2i\sin\frac{\phi}2} \left(u+i(j+\frac12) - \frac12\cot\frac{\phi}{2}\right).
\end{equation}
Here $\phi$ is the twist parameter. If we set
\begin{equation}
\phi=-2\pi i\zeta,\quad j+\frac12 =i(\mu_{\sigma(a)}-x_a),
\end{equation}
then \eqref{na_sum_T11} can be written as a 1-1 component of the product of transfer matrices:
\begin{equation}
\label{eqTprod}
e^{2\pi i\zeta \sum_{a=1}^N x_a}\Big(\prod_{a=1}^N \mathbb{T}^+_{-\frac12 +i(\mu_{\sigma(a)}-x_a)}(u-x_a)\Big)_1^1,
\end{equation}
where $x_a$ are so far undetermined constants (spectral parameters of the Verma modules). This product of matrices is still a diagonal $2\times 2$ matrix, and its only other non-zero component must correspond to $\langle T[u]^2_2\rangle$, which we have not computed yet. Before considering it, let us improve the computation involving $T[u]^1_1$. 

Namely, we can do better, and find the $L$-point function of $T[u]^1_1$ for general $L>1$. The computation of $\langle T[u_1]^1_1 T[u_2]^1_1 \dots T[u_L]^1_1 \rangle$ boils down to the insertion of 
\begin{equation}
\label{L-point}
\prod_{a=1}^N (u_1 - \sigma_a)(u_2-\sigma_a)\dots(u_L-\sigma_a)
\end{equation}
under the contour integral \eqref{CouInt}. Again, this does not change the poles, only modifies their residues. Each $u_j-\sigma_a$ introduces an extra factor of
\begin{equation}
u_j - \mu_{\sigma(a)} - i (n_a + 1/2),
\end{equation}
and the summation over $n_a$'s completely factorizes, so we can separately sum over each $n_a$. Looking at the factor for a given $a$, the expression \eqref{sum_na} suggests that we can replace $n_a$ by $\frac{i}{2\pi}\frac{\dd}{\dd\zeta}$ acting on the $a$'th factor in \eqref{sum_na} (with $e^{2\pi i\zeta\mu_{\sigma(a)}}$ pulled out in front). At the end we find that \eqref{L-point} is equivalent to the following insertion:
\begin{equation}
\label{T11Ltimes}
\prod_{a=1}^N\left[2\sinh(\pi\zeta) \left(u_1 - \mu_{\sigma(a)} + \frac{i}{2\pi}\frac{\dd}{\dd\zeta} \right)\left(u_2 - \mu_{\sigma(a)} + \frac{i}{2\pi}\frac{\dd}{\dd\zeta} \right)\dots \left(u_L - \mu_{\sigma(a)} + \frac{i}{2\pi}\frac{\dd}{\dd\zeta} \right)\frac1{2\sinh(\pi\zeta)} \right].
\end{equation}
Now we compare it to the transfer matrix and Q-operators, again relying on and slightly generalizing the computations from \cite{Bazhanov:2010ts}. They found expressions for the $Q$-operators as traces of certain monodromy matrices over the Fock modules of an auxiliary harmonic oscillator. Their expressions apply to the homogeneous case, that is $u_1=u_2=\dots =u_L$, but are easily generalizable to the case of distinct $u_i$'s. Namely, we write:\footnote{Here we included a somewhat asymmetric and aesthetically unpleasing factor of $e^{\pm\frac12\phi u_1}$ in the definition of $\mathbf{Q}_\pm(\vec{u})$ to generalize the homogeneous case definition of \cite{Bazhanov:2010ts}. Because Q-operators always appear in pairs $\mathbf{Q}_+(\vec{u}-a\vec{e})\mathbf{Q}_-(\vec{u}-b\vec{e})$, we could instead use $e^{\pm\frac12 \phi u_j}$ with some other $j$, or even a more symmetric $e^{\pm\frac1{2L}\phi(u_1+u_2+\dots+u_L)}$, without affecting the transfer matrix. Perhaps it would be more natural not to include this exponential in the definition of $\mathbf{Q}_\pm(\vec{u})$ at all, and rather modify the relation to the transfer matrix \eqref{TeqQQ} as follows:
\begin{equation}
\mathbb{T}^+_{j}(\vec{u}) = \frac{e^{i\phi(j+1/2)}}{2\sinh\pi\zeta} \mathbf{Q}_+(\vec{u} +i(j+1/2)\vec{e})\textbf{Q}_-(\vec{u} -i(j+1/2) \vec{e}).
\end{equation} }
\begin{equation}
\mathbf{Q}_\pm(\vec{u}) = Z^{-1} e^{\pm\frac{1}{2}\phi u_{1}}\tr_{\cF}(\cM_\pm(u_1,\dots, u_L)),\quad Z=\tr_\cF (e^{-i\phi h})=\frac1{2i\sin\frac{\phi}{2}},
\end{equation}
where the trace is over the harmonic oscillator Fock space, the monodromy matrices are:
\begin{equation}
\cM_\pm(u_1,\dots,u_L)=e^{-i\phi h}L_\pm(u_1)\otimes \dots \otimes L_\pm(u_L),
\end{equation}
with 
\begin{equation}
L_+(u)=\left(\begin{matrix}u-ih & ia^+\\ -a^- & 1\end{matrix} \right),\quad L_-(z)=\left( \begin{matrix}1 & a^+\\ ia^- & u + ih\end{matrix} \right),
\end{equation}
where $(a^+, a^-)$ describe the harmonic oscillator, and $h=a^+ a^- + \frac12$ is its Hamiltonian. These Q-operators yield the transfer matrix of the weight-$j$ Verma module:
\begin{equation}
\label{TeqQQ}
\mathbb{T}^+_{j}(\vec{u}) = \frac1{2\sinh\pi\zeta} \mathbf{Q}_+(\vec{u} +i(j+1/2)\vec{e})\textbf{Q}_-(\vec{u} -i(j+1/2) \vec{e}).
\end{equation}
Both Q-operators and the transfer matrix are $2^L\times 2^L$ matrices acting on $(\C^2)^{\otimes L}$. Denoting the natural tensor product basis as $|a_1,\dots,a_L;L\rangle$, where each $a_i=1,2$, we would like to compute the matrix element between $\langle 1,\dots,1;L|$ and $|1,\dots,1;L\rangle$.
Explicit computation of traces over the Fock space shows that $\langle 1,\dots,1;L|\textbf{Q}_-(\vec{u})|1,\dots,1;L\rangle=e^{-\frac{1}{2}\phi u_1}$, and 
\begin{align}
\langle 1,\dots,1;L|&\textbf{Q}_+(\vec{u})|1,\dots,1;L\rangle=e^{\frac{1}{2}\phi u_1}Z^{-1}\tr_\cF e^{-i\phi h} (u_1-ih)(u_2-ih)\dots(u_L-ih)\cr
&=e^{\frac{1}{2}\phi u_1} \frac1{\tr_\cF e^{-i\phi h}} \left(u_1 +\frac{\dd}{\dd\phi}\right)\left(u_2 +\frac{\dd}{\dd\phi}\right)\dots \left(u_L +\frac{\dd}{\dd\phi}\right) \tr_\cF e^{-i\phi h}\cr
&= e^{\frac{1}{2}\phi u_1} 2i\sin\frac{\phi}{2} \left(u_1 +\frac{\dd}{\dd\phi}\right)\left(u_2 +\frac{\dd}{\dd\phi}\right)\dots \left(u_L +\frac{\dd}{\dd\phi}\right) \frac1{2i\sin\frac{\phi}{2}}.
\end{align}
This leads to $\mathbb{T}^+_j$ that clearly agrees with the above expression \eqref{T11Ltimes} upon substitution $\phi=-2\pi i\zeta$, and proper identification of weights and spectral parameters like in \eqref{eqTprod}. We can thus make the following proposal for the correlators in 3d SQCD:
\begin{align}
\label{General_answer_prel}
&u_1^{N}u_2^N\dots u_L^N\langle T[u_1]^{a_1}_{b_1}\dots T[u_L]^{a_L}_{b_L} \rangle \cr
&=\sum_{\sigma\in\substack{\frac{S_{2N}}{S_{N}\times S_N}}} \frac{i^{-N^2} e^{2\pi i\zeta\sum_{a=1}^N x_a} }{\prod_{a=1}^N \prod_{k=N+1}^{2N}2\sinh \pi(\mu_{\sigma(a)}-\mu_{\sigma(k)})} \Big\langle a_1,\dots,a_L;L\Big| \prod_{a=1}^N \mathbb{T}^+_{-\frac12 +i(\mu_{\sigma(a)}-x_a)}(\vec{u}-x_a\vec{e})\Big|b_1,\dots,b_L;L\Big\rangle,\cr
\end{align}
which also includes the case $L=0$ of the partition function \eqref{SQCD_Z} without insertions. It formally corresponds to the length-0 spin chain that has 
\begin{equation}
\label{L0T}
\mathbb{T}^+_j= \frac{e^{-2\pi i \zeta(j+\frac12)}}{2 \sinh \pi\zeta}.
\end{equation}

The answer \eqref{General_answer_prel} still includes the unknown spectral parameters $x_a$. In principle, they are encoded in other correlators that we have not computed yet, such as those of $T[u]^2_2$. The one-point function $\langle T[u]^2_2\rangle$, according to the above expression, can be seen to lead to the following insertion in the sum over vacua:
\begin{equation}
\label{T22_expect}
u^{-N} \prod_a \left\{u + \mu_{\sigma(a)} -2x_a + \frac{i}{2}\coth\pi\zeta\right\},
\end{equation}
computing which would thus be enough to determine all the $x_a$'s.

The expression for the corresponding shift operator is given in \eqref{RTT_sl2}, and can be expanded as
\begin{align}
\label{T22shift}
u^N T[u]^2_2 &=\frac{Q_0(u+\frac{\epsilon}2)Q_2(u+\frac{\epsilon}2)}{Q_1(u+\epsilon)} - \prod_a (u-\phi_a)\sum_a \frac1{(u-\phi_a)(u-\phi_a+\epsilon)}\cdot\frac{Q_0(\phi_a - \frac{\epsilon}{2})Q_2(\phi_a -\frac{\epsilon}2)}{\prod_{b\neq a} (\phi_a-\phi_b)(\phi_a-\phi_b-\epsilon)}\cr &+ \text{ monopole terms},
\end{align}
where the ``monopole terms'' have non-zero magnetic charge and thus vanish in one-point functions. The terms in the first line, despite a cumbersome look, have some nice properties. One can easily check by computing residues that the poles at $\phi_a=\phi_b$ are in fact absent. The poles at $\phi_a = \phi_b + \epsilon$, that is at $\sigma_a=\sigma_b+i$, are present, but are canceled by the sinh-Vandermonde in \eqref{CouInt}. Thus no new poles appear in \eqref{CouInt}, and we still sum over the same poles in the contour integral. Furthermore, one can also check that there are no poles in $u$, so the first line of \eqref{T22shift} is secretly a degree-N polynomial in $u$, which also nicely agrees with the expectation \eqref{T22_expect} for the one-point function.

Still, even with these nice properties, we find it hard to fully compute $\langle T[u]^2_2\rangle$. We can notice, however, that the second piece above (and all monopole terms too) starts contributing at order $u^{-2}$. Therefore computations simplify if we only focus on the $u^{-1}$ order, which we do to test the answer,
\begin{equation}
\langle T[u]^2_2\rangle = \left\langle \frac{Q_0(u+\frac{\epsilon}2)Q_2(u+\frac{\epsilon}2)}{u^N Q_1(u+\epsilon)} \right\rangle + O\left(\frac1{u^2}\right).
\end{equation}
Performing a calculation similar to the one before, we find the insertion of
\begin{equation}
1 + \frac1{u} \sum_a \left\{ \mu_{\sigma(a)} - \mu^L_a - \mu^R_a + \frac{i}{2} \coth\pi\zeta \right\} + O\left(\frac1{u^2}\right)
\end{equation}
in the sum over vacua in \eqref{SQCD_Z}. Comparing with the expectation \eqref{T22_expect}, we see that 
\begin{equation}
\sum_a x_a = \frac12 \sum_a (\mu^L_a + \mu^R_a).
\end{equation}
This determines the spectral parameter in the $N=1$ case, i.e. for the abelian theory:
\begin{equation}
x_1 = \frac12 (\mu^L_1+\mu^R_1).
\end{equation}
Thus for the abelian theory, the two vacua labeled by $\sigma\in S_2$ give the following insertions:
\begin{align}
\label{abel_ans}
\sigma(1)=1:\quad \mathbb{T}^+_{-\frac12 +i\frac{\mu_1-\mu_2}{2}}(\vec{u}-\frac{\mu_1+\mu_2}{2}\vec{e})=\frac1{2\sinh\pi\zeta}\mathbf{Q}_+(\vec{u}-\mu_1 \vec{e})\mathbf{Q}_-(\vec{u}-\mu_2 \vec{e}),\cr
\sigma(1)=2:\quad \mathbb{T}^+_{-\frac12 +i\frac{\mu_2-\mu_1}{2}}(\vec{u}-\frac{\mu_1+\mu_2}{2}\vec{e})=\frac1{2\sinh\pi\zeta}\mathbf{Q}_+(\vec{u}-\mu_2 \vec{e})\mathbf{Q}_-(\vec{u}-\mu_1 \vec{e}).\cr
\end{align}
Essentially, $\mathbf{Q}_+(\vec{u}-\mu_1 \vec{e})\mathbf{Q}_-(\vec{u}-\mu_2 \vec{e})$ and $\mathbf{Q}_+(\vec{u}-\mu_2 \vec{e})\mathbf{Q}_-(\vec{u}-\mu_1 \vec{e})$ give the two traces over the Verma modules of the $N=1$ algebra, and \eqref{General_answer_prel} determines their linear combination that computes sphere correlators \cite{Gaiotto:2019mmf}, which is simply
\begin{equation}
\frac{-i e^{i\pi\zeta(\mu_1 + \mu_2)}}{2\sinh\pi\zeta \,\,2\sinh\pi(\mu_1-\mu_2)}\left(\mathbf{Q}_+(\vec{u}-\mu_1 \vec{e})\mathbf{Q}_-(\vec{u}-\mu_2 \vec{e})-\mathbf{Q}_+(\vec{u}-\mu_2 \vec{e})\mathbf{Q}_-(\vec{u}-\mu_1 \vec{e}) \right).
\end{equation}

What can we do in the $N>1$ case, where the direct computation of $\langle T[u]^2_2 \rangle$ is too hard? The coproduct of the Coulomb branch algebras comes to the rescue. We consider a map
\begin{equation}
\cA_{N;2}[m] \to \cA_{1;2}[m]\otimes \cA_{1;2}[m] \otimes \dots \otimes \cA_{1;2}[m],
\end{equation}
with $N$ factors on the right. Choosing such a map also involves distributing the $2N$ mass parameters of the $U(N)$ theory among the $N$ abelian factors on the right. Each $\cA_{1;2}[m]$ corresponds to a $U(1)$ theory with two massive hypers, and has two traces determined by the Q-operators, as we have just shown. Applying trace in each $\cA_{1;2}[m]$ and multiplying them (as matrices, since every $\mathbf{Q}_\pm$ is a $2^L\times 2^L$ matrix), we are able to generate traces on $\cA_{N;2}[m]$. They are also given by products of Q-operators and have the form:
\begin{equation}
\label{Qprod_tr}
\prod_{a=1}^N \mathbf{Q}_+(\vec{u}-\mu_{\sigma(a)}\vec{e})\mathbf{Q}_-(\vec{u}-\mu_{\sigma(a+N)}\vec{e}),
\end{equation}
where a permutation $\sigma\in S_{2N}$ determines the distribution of $2N$ masses into $N$ pairs $(\mu_{\sigma(a)}, \mu_{\sigma(a+N)})$, $a=1..N$. It also determines the choice of one of the two traces in each factor: $\mathbf{Q}_+(\vec{u}-\mu_{\sigma(a)}\vec{e})\mathbf{Q}_-(\vec{u}-\mu_{\sigma(a+N)}\vec{e})$ as opposed to $\mathbf{Q}_+(\vec{u}-\mu_{\sigma(a+N)}\vec{e})\mathbf{Q}_-(\vec{u}-\mu_{\sigma(a)}\vec{e})$. Because all Q-operators commute with each other, the expression above only depends on $\sigma\in\frac{S_{2N}}{S_N\times S_N}$, so we generate ${2N\choose N}$ different traces in this way, determined by which $N$ masses appear in the $\mathbf{Q}_+$ operators (or equivalently by the choice of massive vacua). They correspond to traces over some modules of the Coulomb branch algebra: as the equation \eqref{General_answer_prel} suggests (because it involves transfer matrices), they are traces over some highest weight modules, which must be quotients of Verma modules. At generic masses, the Verma modules are irreducible, so the traces are simply over the Verma modules themselves. Matching $\sum_a(-\frac12 +i(\mu_{\sigma(a)}-x_a))$ with highest weights of the Verma modules of the truncated Yangian generated by $E(z)$, $F(z)$, $H(z)$ (see Section \ref{sec:Coulomb}), we can even find $x_a$. Note that we have found ${2N\choose N}$ Verma traces, which is a complete set, and any other trace is expressed through them \cite{Etingof:2019guc}.

To solve the problem, we have to find precisely which linear combination of these traces computes the $S^3$ correlators. To this end, we can simply recognize \eqref{General_answer_prel} as such a linear combination, provided that $x_a$ are chosen properly. This really means that we write a general linear combination of traces, and fix its coefficients by looking at the correlators of $T[u]^1_1$. We find that in \eqref{General_answer_prel} we must choose
\begin{equation}
x_a = \frac12 (\mu_{\sigma(a)} + \mu_{\sigma(a+N)}),
\end{equation}
so that writing the answer in terms of Q-operators, we obtain the correct arguments of $\mathbf{Q}_\pm$ as in \eqref{Qprod_tr}. This leads to the answer:
\begin{align}
\label{General_answer}
u_1^{N}u_2^N\dots u_L^N&\langle T[u_1]^{a_1}_{b_1}\dots T[u_L]^{a_L}_{b_L} \rangle \cr
=\sum_{\sigma\in\substack{\frac{S_{2N}}{S_{N}\times S_N}}} &\frac{i^{-N^2} e^{i\pi \zeta\sum_{j=1}^{2N} \mu_j} }{(2\sinh\pi\zeta)^N\prod_{a=1}^N \prod_{k=N+1}^{2N}2\sinh \pi(\mu_{\sigma(a)}-\mu_{\sigma(k)})}\cr
&\times\Big\langle a_1,\dots,a_L;L\Big| \prod_{a=1}^N \mathbf{Q}_+(\vec{u}-\mu_{\sigma(a)}\vec{e})\mathbf{Q}_-(\vec{u}-\mu_{\sigma(a+N)}\vec{e})\Big|b_1,\dots,b_L;L\Big\rangle,\cr
\end{align}

Next we couple it to the 4d bulk on the two sides of the interface by integrating over masses. We do a different analysis here than in Section \ref{sec:XXX}. It is convenient to go back to the integral expression for the 3d partition function (with an insertion of $T[u_1]^1_1\dots T[u_L]^1_1$) and write:
\begin{align}
\int \frac{[\dd\mu][\dd\nu][\dd\sigma]}{(N!)^3}e^{-\frac{i\pi}{\tau}\tr \mu^2 - \frac{i\pi}{\tau}\tr \nu^2}\Delta(\mu)\Delta(\nu) \frac{\mathbbm{\Delta}(\mu)\mathbbm{\Delta}(\sigma)}{\prod_{a,b=1}^N 2\cosh\pi(\sigma_a - \mu_b)}  \frac{\mathbbm{\Delta}(\nu)\mathbbm{\Delta}(\sigma)}{\prod_{a,b=1}^N 2\cosh\pi(\sigma_a - \nu_b)}\cr \times\prod_{j=1}^N \left[e^{2\pi i \zeta\sigma_j} \prod_{i=1}^L(u_i - \sigma_j) \right].
\end{align}
Now we apply the Cauchy identity to the ratios appearing in the above formula:
\begin{align}
\frac{\mathbbm{\Delta}(\mu)\mathbbm{\Delta}(\sigma)}{\prod_{a,b=1}^N 2\cosh\pi(\sigma_a - \mu_b)}=\sum_{s\in S_N} (-1)^s \prod_{a=1}^N \frac1{2\cosh\pi(\sigma_a - \mu_{s(a)})},
\end{align}
and the same for the second one. Using the Weyl symmetry, this cancels two factors of $N!$ and we obtain
\begin{align}
\frac1{N!}\int[\dd \mu][\dd \nu][\dd\sigma] e^{-\frac{i\pi}{\tau}\tr \mu^2 - \frac{i\pi}{\tau}\tr \nu^2}\Delta(\mu)\Delta(\nu)\prod_{a=1}^N \frac{e^{2\pi i\zeta\sigma_a}\prod_{i=1}^L(u_i - \sigma_a)}{2\cosh\pi(\sigma_a-\mu_a)\, 2\cosh\pi(\sigma_a-\nu_a)}.
\end{align}
At this point, we can integrate over $\sigma_a$'s by closing the contour and picking up the poles at
\begin{equation}
\sigma_a = \mu_a + i\left(n_a + \frac12\right),\quad \text{and } \sigma_a = \nu_a + i\left(n_a + \frac12 \right),\ n_a\in\Z_{\geq 0},
\end{equation}
where we assumed that $\zeta>0$. Upon further summation over $n_a$, the part of the integrand that involves a product over $a$ can be written as:
\begin{equation}
\prod_{a=1}^N \frac1{2i\sinh\pi(\mu_a-\nu_a)} \left\{ e^{2\pi i\zeta \mu_a}\prod_{k=1}^L\left(u_k-\mu_a-\frac1{2\pi i}\frac{\dd}{\dd\zeta}\right) - e^{2\pi i\zeta \nu_a}\prod_{k=1}^L\left(u_k-\nu_a-\frac1{2\pi i}\frac{\dd}{\dd\zeta}\right) \right\}\frac1{2\sinh\pi\zeta}.
\end{equation}
We recognize an already familiar expression for the transfer matrix components, which allows to write this as
\begin{equation}
\langle 1,\dots,1;L|\prod_{a=1}^N \frac{\mathbb{T}^+_{-\frac12 +i(\mu_a-x_a)}(\vec{u}-x_a\vec{e})-\mathbb{T}^+_{-\frac12 +i(\nu_a-x_a)}(\vec{u}-x_a\vec{e})}{2i\sinh\pi(\mu_a-\nu_a)}|1,\dots,1;L\rangle,
\end{equation}
where the spectral parameters $x_a$ follow from $\langle T[u]^2_2\rangle$, which can be found by comparison with the earlier analysis, giving:
\begin{equation}
x_a=\frac12 (\mu_a + \nu_a)\equiv \bar\mu_a.
\end{equation}
We therefore conclude that on the interface, the correlators are written as
\begin{align}
u_1^{N}\dots u_L^N&\langle T[u_1]^{a_1}_{b_1}\dots T[u_L]^{a_L}_{b_L} \rangle\cr
=\frac1{N!}\int&[\dd \mu][\dd \nu] e^{-\frac{i\pi}{\tau}\tr\mu^2 - \frac{i\pi}{\tau}\tr\nu^2}\Delta(\mu)\Delta(\nu)e^{2\pi i\zeta \sum_{j=1}^N \bar{\mu}_j}\cr 
&\times\Big\langle a_1,\dots,a_L;L\Big|\prod_{j=1}^N \frac{\mathbb{T}^+_{-\frac12 +i(\mu_j-\bar{\mu}_j)}(\vec{u}-\bar{\mu}_j\vec{e})-\mathbb{T}^+_{-\frac12 +i(\nu_j-\bar{\mu}_j)}(\vec{u}-\bar{\mu}_j\vec{e})}{2i\sinh\pi(\mu_j-\nu_j)}\Big|b_1,\dots,b_L;L\Big\rangle,\cr
\end{align}
where the integral is taken over $\R^N\times \R^N$, i.e. the Cartan of $U(N)\times U(N)$. This answer for correlators also includes the partition function with no insertions, and the corresponding $L=0$ transfer matrix is given in \eqref{L0T}.

\section{Identity for Q-operators}\label{app:comb}
Let us prove
\begin{align}
\label{comb_id_copy}
\sum_{\sigma\in\frac{S_{2N}}{S_N\times S_N}} \frac{\mathbbm{\Delta}(\mu^L)\mathbbm{\Delta}(\mu^R)}{\prod_{a=1}^N \prod_{k=N+1}^{2N}2\sinh \pi(\mu_{\sigma(a)}-\mu_{\sigma(k)})} \prod_{a=1}^N \mathbf{Q}_+(\vec{u}-\mu_{\sigma(a)}\vec{e})\mathbf{Q}_-(\vec{u}-\mu_{\sigma(a+N)}\vec{e})\cr
=(-1)^{\frac{N(N-1)}{2}}\sum_{s\in S_N}(-1)^s \prod_{a=1}^N \frac{\mathbf{Q}_+(\vec{u}-\mu^L_a\vec{e})\mathbf{Q}_-(\vec{u}-\mu^R_{s(a)}\vec{e})-\mathbf{Q}_-(\vec{u}-\mu^L_a\vec{e})\mathbf{Q}_+(\vec{u}-\mu^R_{s(a)}\vec{e})}{2\sinh\pi(\mu^L_a-\mu^R_{s(a)})}.
\end{align}
Because Q-operators commute, their order does not matter, what matters is which $N$ out of $2N$ masses appear inside $\mathbf{Q}_+$ (the rest appear in $\mathbf{Q}_-$). This is determined by the permutation $\sigma\in S_{2N}$, so fix $\sigma$, and look at the coefficient of $\prod_{a=1}^N \mathbf{Q}_+(\vec{u}-\mu_{\sigma(a)}\vec{e})\mathbf{Q}_-(\vec{u}-\mu_{\sigma(a+N)}\vec{e})$ on both sides of the equality. Its coefficient on the left is what we see in \eqref{comb_id_copy}:
\begin{equation}
\label{lhs_coeff}
\frac{\mathbbm{\Delta}(\mu^L)\mathbbm{\Delta}(\mu^R)}{\prod_{a=1}^N \prod_{k=N+1}^{2N}2\sinh \pi(\mu_{\sigma(a)}-\mu_{\sigma(k)})}.
\end{equation}
We want to show that we get the same coefficient on the right. Suppose that $(\mu_{\sigma(1)},\dots,\mu_{\sigma(N)})$ contains $p$ left masses $(\mu^L_{\cL_1},\dots,\mu^L_{\cL_p})$ and $N-p$ right masses $(\mu^R_{\cR_1},\dots,\mu^R_{\cR_{N-p}})$. They are indexed by $\cL\subset \{1,\dots,N\}$ and $\cR\subset \{1,\dots,N\}$, and in what follows we call $\cI_N=\{1,\dots,N\}$. To pick only the terms proportional to $\prod_{a=1}^N \mathbf{Q}_+(\vec{u}-\mu_{\sigma(a)}\vec{e})\mathbf{Q}_-(\vec{u}-\mu_{\sigma(a+N)}\vec{e})$ on the right of \eqref{comb_id_copy}, we must consider only those permutations $s\in S_N$, for which
\begin{equation}
\label{s1s2_perm}
s(\cL) = \cI_N\setminus \cR,
\end{equation}
and in the product over $a=1..N$, we pick $\mathbf{Q}_+(\vec{u}-\mu^L_a\vec{e})\mathbf{Q}_-(\vec{u}-\mu^R_{s(a)}\vec{e})$ for those factors with $a\in\cL$, and $\mathbf{Q}_-(\vec{u}-\mu^L_a\vec{e})\mathbf{Q}_+(\vec{u}-\mu^R_{s(a)}\vec{e})$ for those with $a\in\cI_N\setminus\cL$.
Permutations \eqref{s1s2_perm} can be described by picking a reference permutation $s_0$ obeying this property, from which all others are obtained by composing with pairs $(s_1, s_2)\in S_p\times S_{N-p}$, where $s_1: \cL \to \cI_N\setminus \cR$ and $s_2: \cI_N\setminus \cL \to \cR$.

We notice that in \eqref{lhs_coeff}, some $\sinh$'s in the denominator are canceled against the similar ones in the numerator. The ones that remain in the denominator correspond to pairings in $\cL\times (\cI_N\setminus\cR)$ and in $(\cI_N\setminus\cL)\times \cR$. The ones that remain in the numerator are self-parings in the sets $\cL$, $\cI_N\setminus\cL$, $\cR$, and $\cI_N\setminus\cR$. In the end, \eqref{lhs_coeff} factorizes (up to a sign) as
\begin{equation}
\frac{\mathbbm{\Delta}(\mu^L\in\cL)\mathbbm{\Delta}(\mu^R\in\cI_N\setminus\cR)}{\prod_{a\in\cL}\prod_{b\in \cI_N\setminus \cR} 2\sinh\pi(\mu_a-\mu_{N+b})} \times \frac{\mathbbm{\Delta}(\mu^L\in\cI_N\setminus\cL)\mathbbm{\Delta}(\mu^R\in\cR)}{\prod_{a\in\cI_N\setminus\cL}\prod_{b\in \cR} 2\sinh\pi(\mu_{N+b}-\mu_a)}.
\end{equation}
Apply the following version of the Cauchy identity to each of the two factors here,
\begin{equation}
\frac{\mathbbm{\Delta}(x)\mathbbm{\Delta}(y)}{\prod_{z,b=1}^n 2\sinh\pi(x_a-y_b)}= (-1)^{\frac{n(n-1)}{2}}\sum_{s\in S_n} (-1)^s \prod_{a=1}^n \frac{1}{2\sinh\pi(x_a-y_{s(b)})}.
\end{equation}
As a result, we find a sum over $(s_1, s_2)\in S_p\times S_{N-p}$, which precisely matches the corresponding coefficient of Q-operators on the right. Keeping track of signs gives \eqref{comb_id_copy}.$\square$

\setlength{\unitlength}{1mm}

\newpage

\bibliographystyle{utphys}
\bibliography{traces}
\end{document}